\documentclass[aps,prb,twocolumn,10pt,superscriptaddress,notitlepage]{revtex4-1}
\pdfoutput=1
\usepackage[font=sf, labelfont={sf,bf},size=small,justification=RaggedRight]{caption}

\usepackage[utf8]{inputenc}
\usepackage{graphicx}
\usepackage{amsmath}
\usepackage{float}
\usepackage[unicode=true,
	    colorlinks=true,
	    linkcolor=blue,
	    citecolor=blue,
	    urlcolor=blue]{hyperref} 
\usepackage{siunitx}
\usepackage{subfigure}
\usepackage{shorttoc}

\newcommand{\pwl}{\ensuremath{\lambda_{\mathrm{p}}}}
\newcommand{\pq}{\ensuremath{q_{\mathrm{p}}}}
\newcommand{\pinvloss}{\ensuremath{\gamma_{\mathrm{p}}^{-1}}}

\renewcommand{\Re}{\operatorname{Re}}
\renewcommand{\Im}{\operatorname{Im}}

\newcommand{\tipradius}{\ensuremath{R}}

\newcommand{\sopt}{\ensuremath{\xi_{\mathrm{opt}}}}   
\newcommand{\sbulk}{\ensuremath{\xi_{\mathrm{bulk}}}} 
\newcommand{\ssmooth}{\ensuremath{\xi_{\mathrm{smooth}}}}  
\newcommand{\sedge}{\ensuremath{\xi_{\mathrm{edge}}}}  
\newcommand{\sdiff}{\ensuremath{\delta \xi_{\mathrm{opt}}}}  

\begin{document}
\title{Highly confined low-loss plasmons in graphene--boron nitride heterostructures}

\author{Achim Woessner}
\thanks{These authors contributed equally}
\affiliation{ICFO – The Insititute of Photonic Sciences, Mediterranean Technology Park, 08860 Castelldefels (Barcelona), Spain}
\author{Mark B. Lundeberg}
\thanks{These authors contributed equally}
\affiliation{ICFO – The Insititute of Photonic Sciences, Mediterranean Technology Park, 08860 Castelldefels (Barcelona), Spain}
\author{Yuanda Gao}
\thanks{These authors contributed equally}
\affiliation{Department of Mechanical Engineering, Columbia University, New York, NY 10027, USA}
\author{Alessandro Principi}
\affiliation{Department of Physics and Astronomy, University of Missouri, Columbia, Missouri 65211, USA}
\author{Pablo Alonso-González}
\affiliation{CIC nanoGUNE Consolider, 20018 Donostia-San Sebastián, Spain}
\author{Matteo Carrega}
\affiliation{NEST, Istituto Nanoscienze - CNR and Scuola Normale Superiore, 56126 Pisa, Italy}
\affiliation{SPIN-CNR, Via Dodecaneso 33, 16146 Genova, Italy}
\author{Kenji Watanabe}
\affiliation{National Institute for Materials Science, 1-1 Namiki, Tsukuba 305-0044, Japan}
\author{Takashi Taniguchi}
\affiliation{National Institute for Materials Science, 1-1 Namiki, Tsukuba 305-0044, Japan}
\author{Giovanni Vignale}
\affiliation{Department of Physics and Astronomy, University of Missouri, Columbia, Missouri 65211, USA}
\author{Marco Polini}
\affiliation{NEST, Istituto Nanoscienze - CNR and Scuola Normale Superiore, 56126 Pisa, Italy}
\author{James Hone}
\affiliation{Department of Mechanical Engineering, Columbia University, New York, NY 10027, USA}
\author{Rainer Hillenbrand}
\affiliation{CIC nanoGUNE Consolider, 20018 Donostia-San Sebastián, Spain}
\affiliation{IKERBASQUE, Basque Foundation for Science, 48011 Bilbao, Spain}
\author{Frank H.L. Koppens}
\email{frank.koppens@icfo.es}
\affiliation{ICFO – The Insititute of Photonic Sciences, Mediterranean Technology Park, 08860 Castelldefels (Barcelona), Spain}

\maketitle
\textbf{
Graphene plasmons were predicted to possess ultra-strong field confinement and very low damping at the same time, enabling new classes of devices for deep subwavelength metamaterials\cite{Fang2014a,Grigorenko2012}, single-photon nonlinearities,\cite{Gullans2013} extraordinarily strong light-matter interactions,\cite{Koppens_Nano_Lett_2011} and nano-optoelectronic switches.
While all of these great prospects require low damping, thus far strong plasmon damping was observed,\cite{Fei2012,Chen2012,Alonso-Gonzalez2014} with both impurity scattering\cite{Principi2013c} and many-body effects in graphene\cite{Fei2012} proposed as possible explanations.
With the advent of van der Waals heterostructures,\cite{Dean2010,Geim2013} new methods have been developed to integrate graphene with other atomically flat materials.
In this letter we exploit near-field microscopy to image propagating plasmons in high quality graphene encapsulated between two films of hexagonal boron nitride (h-BN).\cite{Wang2013}
We determine dispersion and particularly plasmon damping in real space.
We find unprecedented low plasmon damping combined with strong field confinement, and identify the main damping channels as intrinsic thermal phonons in the graphene and dielectric losses in the h-BN.
The observation and in-depth understanding of low plasmon damping is the key for the development of graphene nano-photonic and nano-optoelectronic devices.
}

Many interesting extraordinary optical and electronic phenomena can occur in graphene--h-BN heterostructures, for example an altered electronic massless Dirac fermion spectrum of graphene\cite{Yankowitz2012} that is predicted to cause plasmon "morphing" with additional satellite plasmon modes.\cite{Tomadin2014}
Furthermore, h-BN itself is an interesting optical material as it is a natural hyperbolic material, supporting tunable propagating phonon polaritons in the bulk.\cite{Dai2014b,Caldwell2014} 
Combining h-BN with graphene gives rise to unconventional plasmon-phonon hybridization\cite{Brar2014} and this hybrid system can be used for tailoring novel subwavelength metamaterials.

Besides all of those exotic properties, h-BN can provide an exceptionally clean environment for graphene. 
Recent advances in graphene device fabrication, exploiting the unique properties of h-BN heterostructures produced by the polymer-free van der Waals assembling technique, resulted in significantly less disorder.
This leads to the carrier transport mobility at room temperature reaching its intrinsic limit dominated by thermal phonon scattering.\cite{Wang2013} 

\begin{figure}[t]
\centering
\includegraphics{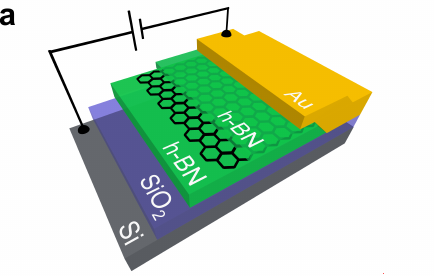}%
\includegraphics{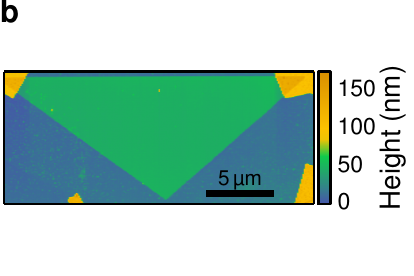}
\includegraphics{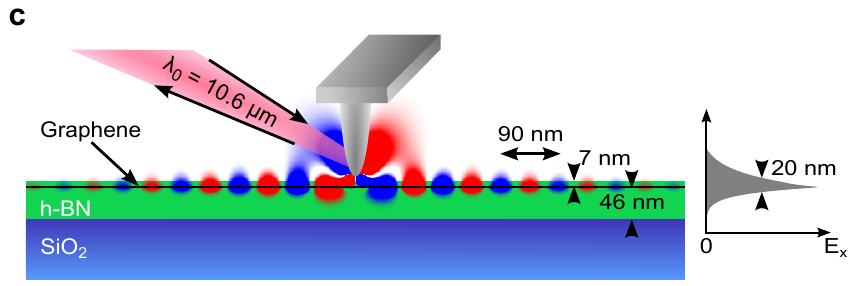}
\includegraphics{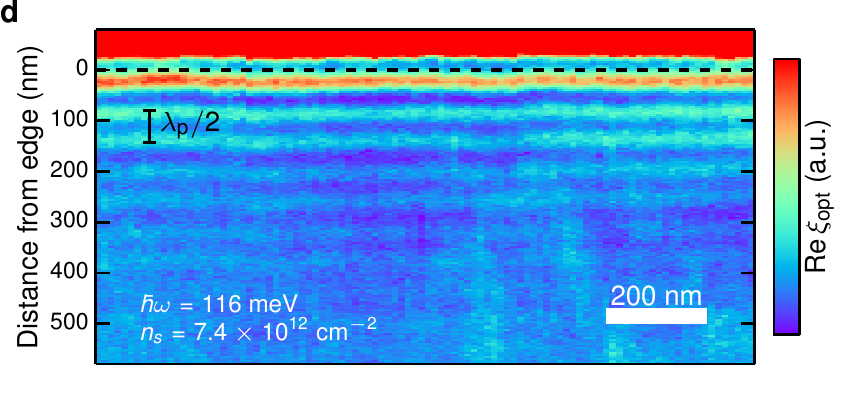}
\caption{
\label{fig1}
\textbf{Device and plasmon imaging with s-SNOM.}
\textbf{a},
Sketch of the layered heterostructure with a Si backgate, SiO$_2$ layer, h-BN, graphene, h-BN, and a gold side contact.
\textbf{b},
Topography image of the device.
The triangle is two h-BN layers encapsulating a graphene layer, contacted at two corners.
The blue outer area is etched.
\textbf{c},
Simplified side-view schematic of the s-SNOM measurement including probe tip, excitation, and detection.
Plasmons are launched radially from the tip.
The color map shows the simulated in-plane component of the electric field of a dipole source oscillating at a photon energy of 116~meV coupling to graphene plasmons.
The simulated field confinement of the plasmon in the out-of-plane direction of 20~nm at full width half maximum can be seen on the right.
\textbf{d},
s-SNOM optical signal from two-dimensional scan of tip position, near the graphene edge at room temperature (dashed line).
Edge-reflected plasmons appear as interference fringes.
}
\end{figure}

Here we exploit this new type of heterostructure, sketched in Fig.~\ref{fig1}a, and show an unprecedented low damping and strong field confinement of graphene plasmons.
Furthermore, we establish an excellent understanding of the graphene plasmon dispersion and damping for a wide range of carrier densities.
In contrast to earlier reports, we find much lower plasmon damping and that impurity scattering does not play a significant role in plasmon damping, pointing at very low intrinsic limits on the plasmon damping in graphene.
This shows that graphene encapsulated in h-BN provides an excellent platform for graphene plasmonic devices.

A topography image of the device is depicted in Fig.~\ref{fig1}b.
The h-BN(7~nm)--graphene--h-BN(46~nm) stack assembled by the polymer-free van der Waals assembling technique\cite{Wang2013} lies on top of an oxidized silicon wafer, used as a backgate.
This stack is etched into a triangle and is electrically side-contacted with metal electrodes.\cite{Wang2013}

We image propagating plasmons with a scattering-type scanning near-field optical microscope (s-SNOM), similar to several recent studies of graphene plasmons.\cite{Chen2012,Fei2012,Alonso-Gonzalez2014} 
A schematic of the s-SNOM interacting with the graphene device is shown in Fig.~\ref{fig1}c.
A continuous wave laser, with tunable photon energy from 115 to 135~meV, is focussed on a metallized atomic force microscope probe tip.
The tip apex optically couples to the device in the near-field.
The sharpness of the apex provides wavevector matching between plasmon and incident photon.\cite{Fei2011}
The incident light is partly converted to plasmons, which propagate away from the tip as a circular wave with complex wavevector $\pq$.
Plasmons return to the tip if they are reflected by edges or defects.
Returning plasmons are partly converted to light and add to the out-scattered light field.
Interferometric detection of the scattered light yields magnitude and phase as the complex-valued optical signal $\sopt$.
A scan of $\Re\sopt$ vs.\ tip position near the graphene edge shows characteristic fringes due to the varying field of the reflected plasmon, interfering with the local response.\cite{Chen2012,Fei2012}
Figure~\ref{fig1}c shows these fringes measured near a straight edge in our device.
Since the plasmon returns to the tip after travelling twice the tip-edge distance, the spacing between fringes is $\pwl/2$, where $\pwl = 2\pi/\Re\pq$ is the plasmon wavelength.\cite{Chen2012,Fei2012}

\begin{figure}[t]
\begin{minipage}{0.5\columnwidth}
\includegraphics{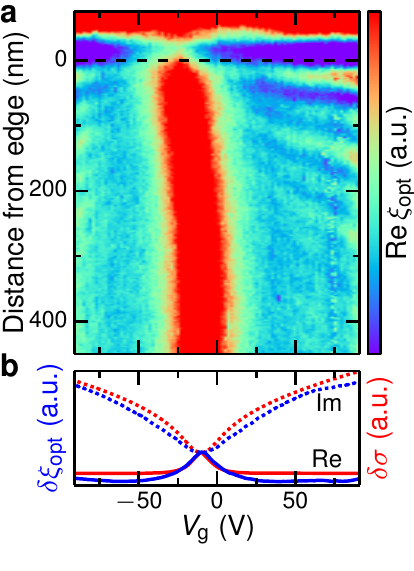}
\includegraphics{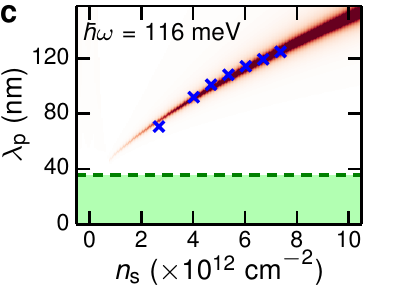}
\end{minipage}\hfill%
\begin{minipage}{0.5\columnwidth}
\includegraphics{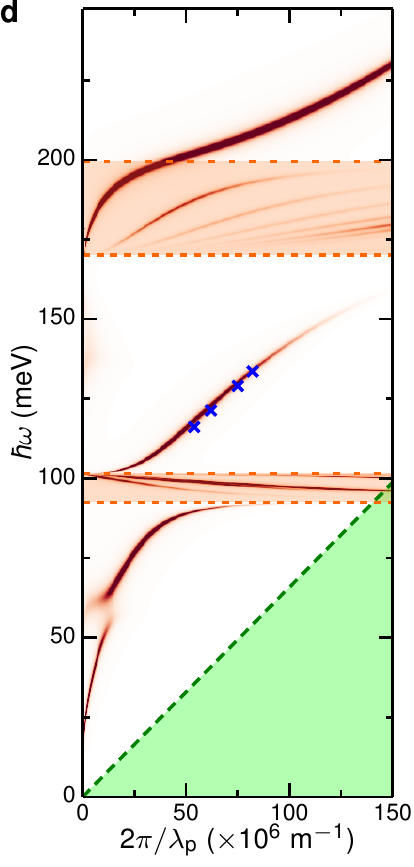}
\end{minipage}
\caption{
\label{fig2}
\textbf{Optical signal and plasmon wavelength dependence on carrier density and photon energy.}
\textbf{a},
s-SNOM optical signal from scan of tip position perpendicular to the graphene edge (dashed line) and gate voltage,
showing gate-dependence of plasmon fringes at a photon energy of $\hbar\omega=116~\mathrm{meV}$.
\textbf{b},
Change in complex optical signal away from the edge (blue, from \textbf{a}) with respect to gate voltage, compared to theoretical local conductivity for ideal graphene (red).\cite{Koppens_Nano_Lett_2011}
\textbf{c},
Plasmon wavelength dependence on carrier density.
\textbf{d},
Dependence on frequency, at $n_\mathrm{s} = 7.4\times10^{12}~\mathrm{cm^{-2}}$.
Shaded orange regions indicate the h-BN frequency bands in which propagating phonon polaritons can exist. 
In both \textbf{c} and \textbf{d}, crosses show the extracted experimental values and the red background color plot shows the imaginary part of the Fresnel reflection coefficient (see Methods).
The electronic intraband Landau damping region is shaded green.
}
\end{figure}

Due to the encapsulation of the graphene, our device possesses only small intrinsic doping and a uniform doping distribution with a small density of electron-hole puddles.\cite{Xue2011}
This enables us to study the optical response for a wide range of carrier densities $n_\mathrm{s}$, including features near the charge neutrality point, by applying a backgate voltage $V_\mathrm{g}$.
In Fig.~\ref{fig2}a we tune the plasmon fringes in both wavelength and amplitude and show that $\pwl$ depends strongly on $n_\mathrm{s}$.
With decreasing carrier density the fringe visibility decreases, as the wavelength of plasmons becomes shorter.
The tip cannot couple to plasmons with an arbitrarily short wavelength due to the non-zero tip radius\cite{Fei2011} and their confinement in the top h-BN layer.

While changing $n_\mathrm{s}$ we also observe changes in the local optical response.
This is most clearly seen in Fig.~\ref{fig2}b where we plot $\sopt$ versus $n_\mathrm{s}$ with the signal averaged from 400~nm to 700~nm from the edge, where plasmon interference effects are weak.  
With appropriately chosen phase, $\sopt$ is approximately proportional to the change in complex valued graphene conductivity $\sigma$ (Fig.~\ref{fig2}b, see Supplement).
Near charge neutrality (small $|n_\mathrm{s}|$), $\Re\sigma$ dominates which gives information about interband conductivity.
A corresponding peak in $\Re\sopt$ appears where graphene is charge neutral, in this case near $V_\mathrm{g} \approx -10$~V.
With increasing carrier density $\Re\sigma$ ($\Re\sopt$) decreases due to Pauli blocking and $\Im\sigma$ ($\Im\sopt$) grows due to ballistic free carrier motion (Drude-like response).
With this technique we confirm the spatial uniformity of the position of the graphene charge neutrality point and deduce that plasmons are hosted in graphene with uniform carrier density.

A detailed study of the plasmon wavelength dependence on carrier density and frequency is shown in Fig.~\ref{fig2}c,d together with calculations of the graphene plasmon dispersion of the full system.
The calculations include optical thin film effects which need to be included due to the thin h-BN top film as well as graphene nonlocal conductivity which needs to be considered due to the low plasmon phase velocity (see Supplement).
The measured wavelengths show parameter-free agreement with these electromagnetic calculations (red curves in Fig.~\ref{fig2}c,d; details in Methods).
The additional modes (in the orange bands in Fig.~\ref{fig2}d) that appear in the calculation are due to the propagating phonon polaritons in thin h-BN.\cite{Dai2014b,Caldwell2014}
These phonon modes can hybridize with the graphene plasmons, however the plasmons are effectively unhybridized for the frequency range used in this study.

For our frequency range, the h-BN lattice is non-resonant yet yields a highly anisotropic dielectric environment for the plasmon, which enhances its confinement. 
The out-of-plane full width at half maximum confinement of the plasmon electric field is calculated to be $\sim 20$~nm (Fig.~\ref{fig1}c).  
We observe $\pwl$ as low as $70~\mathrm{nm}$, 150 times smaller than the free space light wavelength. 
This constitutes a record high volume confinement of propagating optical fields of $\sim 10^7$ compared to the modal volume in free space.

\begin{figure}[t]
\includegraphics{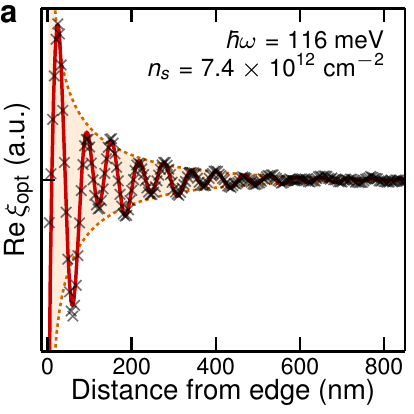}
\includegraphics{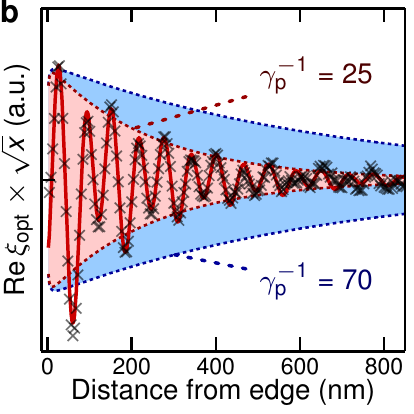}
\caption{
\label{fig3}
\textbf{Extraction of plasmon damping.}
\textbf{a},
Black crosses show the s-SNOM optical signal from Fig.~\ref{fig1}d, averaged along the edge, with a smooth background subtracted.
A fit to equation~\eqref{eq:fit} is shown as a red line.
The shaded region shows the decay envelope of the first term of equation \eqref{eq:fit}.
\textbf{b},
Same signal and fit as in \textbf{a}, multiplied by $\sqrt{x}$ to isolate exponential decay.
The shaded regions show the exponential decay envelopes of the first term of equation \eqref{eq:fit} for the measured damping ($\pinvloss = 25$) and for the case limited only by electron scattering from thermal phonons ($\pinvloss = 70$).
}
\end{figure}

The capability to carry plasmons with such strong field confinement and at the same time relatively low propagation damping is a unique property of graphene compared to other plasmonic materials.\cite{Jablan_PRB_2009}  
In order to quantify the propagation damping, we average linescans of the complex $\sopt$ perpendicular to the graphene edge at different locations, and subtract the background (Fig.~\ref{fig3}a). 
The decay of the fringes away from an edge is due to a combination of damping ($\Im\pq > 0$) and circular-wave geometrical spreading. 
The oscillating signal of Fig.~\ref{fig3}a fits well with:
\begin{equation}
\sopt(x) = A\dfrac{e^{i2\pq x}}{\sqrt{x}} + B \dfrac{e^{i\pq x}}{x^{a}},
\label{eq:fit}
\end{equation}
with complex parameters $A$, $B$, $\pq$ and real $a$.
The first term is the returning field for a damped circular wave reflected from a straight edge, with the plasmon travelling $2x$.
The second term interferes with the first, producing alternating fringe amplitudes.
It arises because plasmons are not only generated/detected beneath the tip apex, but also weakly at the edge of the graphene.\cite{Zhang2014a}
These plasmons travel only the tip-edge distance $x$ and therefore show twice the fringe spacing of the plasmons generated/detected beneath the tip apex.
As the geometrical decay of the plasmon travelling the tip-edge distance only once is not known \textit{a priori}, we allow for a variable decay $a \sim 1$.
Nevertheless, because the $\exp(2i\pq x)$ component dominates and we can separate the $\exp(i\pq x)$ component with Fourier analysis, we can extract $\Im\pq$ unambiguously (see Supplement).

We define the inverse damping ratio $\pinvloss=\Re\pq/\Im\pq$ as dimensionless figure of merit of propagation damping.
Fig.~\ref{fig3}b shows the data multiplied with $\sqrt{x}$ to isolate the damping decay $\exp(-2\Im \pq x)$, and visually indicates the significance of $\pinvloss$---in this case, $\pinvloss \approx 25$.
This is a significant improvement over the $\pinvloss \sim 5$ seen in studies of unencapsulated graphene on silicon dioxide.\cite{Chen2012,Fei2012}

From spatial damping, the plasmon amplitude decay time $\tau_\mathrm{p}$ can be calculated using the group velocity $v_\mathrm{g}=(d\Re\pq / d\omega)^{-1}$.
In this case $v_\mathrm{g} \approx 10^6~\mathrm{m/s}$, coincidentally the same as the Fermi velocity of graphene electrons (see Fig.~\ref{fig2}d).
We find $\tau_\mathrm{p} = (\Im\pq)^{-1}/v_\mathrm{g} \approx 500~\mathrm{fs}$, which is remarkably long for strongly confined optical fields, and an order of magnitude longer than the 
amplitude decay time of subwavelength plasmons in silver, the metal with the longest plasmon amplitude decay time.\cite{Johnson1972}

\begin{figure}[t]
\includegraphics{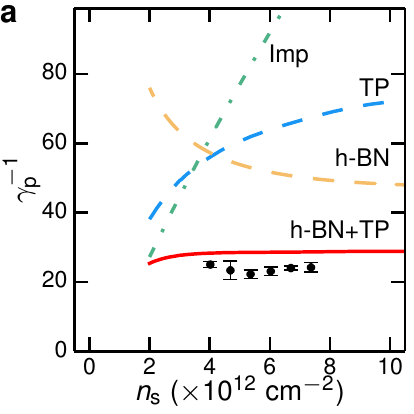}
\includegraphics{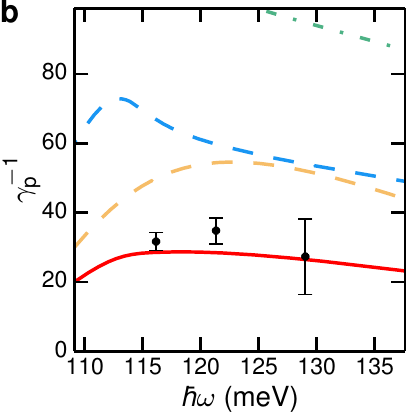}
\caption{
\label{fig4}
\textbf{Plasmon damping mechanisms.}
\textbf{a},
The inverse damping ratio as a function of carrier density at a photon energy of 116~meV.
\textbf{b},
The inverse damping ratio as a function of excitation frequency at a carrier density of $7.4\times10^{12}~\mathrm{cm^{-2}}$.
Both \textbf{a} and \textbf{b} also show the theoretical inverse damping ratios due to
 graphene thermal phonons (blue dashed curve),
 hypothetical charge impurities at concentration $n_{\rm imp} = 1.9\times 10^{11}~\mathrm{cm^{-2}}$ (green dash-dotted curve),
 dielectric losses of h-BN (yellow dashed curve)
 and the combination of graphene thermal phonons and dielectric losses of h-BN (red curve).
}
\end{figure}

Plasmon damping can arise from a number of mechanisms, most of which involve electron scattering.
Electrons can  scatter from the disorder potential in the graphene, created by extrinsic charge impurities\cite{Principi2013c} and the intrinsic thermal phonons.\cite{Principi}
Electrons may also inelastically scatter by absorbing energy from the plasmon while emitting an optical phonon in the graphene or in the substrate.\cite{Yan2013c,Principi}
Coherent many-electron scattering processes have been calculated to play a minor role.\cite{Principi2013b}
Besides dissipating energy electronically, plasmons also dissipate via dielectric losses in the environment.\cite{Principi2013c,Principi}

We investigate the role of the different damping mechanisms by measuring the inverse damping ratio as a function of both $n_{\rm s}$ and excitation frequency, and compare the results in Fig.~\ref{fig4} with the calculated damping for various damping channels. 
The calculations are based on the nonlocal conductivity $\sigma(q,\omega)$ evaluated  at the plasmon wave vector $q = q_{\rm p}$ and at the excitation frequency $\omega$. 
These calculations show that charge carrier scattering is strongly modified at high frequency and thus the effective electron scattering time for plasmons can differ from the transport scattering time.\cite{Principi2013c,Principi} 

Interestingly, we find experimentally that plasmon damping is not affected by the carrier density. 
By comparing our data with the calculated inverse damping ratios, we find that impurity scattering does not play a role for plasmon damping, because it would lead to strongly reduced damping for increasing $n_{\rm s}$, due to increasing electrostatic screening\cite{Principi2013c} (green dashed dotted curve in Fig.~\ref{fig4}). 
In contrast, plasmon damping by intrinsic thermal phonons\cite{Principi} shows a much weaker dependence on $n_{\rm s}$ (dashed blue curve in Fig.~\ref{fig4}).
From quantitative comparison, we find that (without fitting parameters) this intrinsic damping mechanism accounts for approximately half the observed damping and thus we conclude that this is the dominant intrinsic damping mechanism. 
Extensive details on the calculations of plasmon damping due to thermal phonon scattering are presented in Ref.~\onlinecite{Principi}. 

Electronic damping alone cannot explain the observed dependences, however dielectric losses provide an additional damping pathway.
In particular, the dielectric losses of the h-BN encapsulating the graphene may give a significant contribution (yellow dashed curve in Fig.~\ref{fig4}).
The combination of thermal phonon damping and dielectric losses\cite{Principi} is in good agreement with our measurements (red curve in Fig.~\ref{fig4}).
The dielectric losses used in our model are consistent with recent measurements of thin ($<$200~nm) h-BN flakes.\cite{Caldwell2014} 
Although the plasmon damping is affected by the dielectric losses, this work provides strong evidence of the intrinsic limit graphene plasmon inverse damping ratio of 40--70. 
This provides a upper bound on $\Re\sigma\sim0.05\pi e^2/2h$ at room temperature, much smaller than previously reported.\cite{Li2008f,Mak2008b}

To conclude,
we have demonstrated h-BN to be an exceptional environment for graphene plasmons, yielding high confinement and low levels of damping.
In order to further reduce damping and reach the ultimate limit of plasmon propagation at room temperature---electron scattering by thermal phonons\cite{Principi}---it will be necessary to reduce dielectric losses.
The presented nano-photonics device paves the way towards single-photon nonlinearities with graphene plasmons,\cite{Gullans2013} and provide an ideal platform for many applications where tunability is crucial, such as routing of plasmons\cite{Christensen2012} and plasmon lenses.\cite{Alonso-Gonzalez2014,Vakil2011}

\section*{Methods}
{\small

The device geometry as well as the edge contacts were defined using electron beam lithography and dry etching, in the method of Ref.~\onlinecite{Wang2013}.
The backgate capacitance density was estimated to be $6.7\times 10^{10}~e\,\mathrm{cm^{-2}\,V^{-1}}$, where $e$ is the elementary charge.

The s-SNOM used was a NeaSNOM from Neaspec GmbH, equipped with a CO$_2$ laser and cryogenic HgCdTe detector.
The probes were commercially-available metallized atomic force microscopy probes with an apex radius of approximately 25~nm.
The tip height was modulated at a frequency of approximately $250~\mathrm{kHz}$ with amplitude of 60--80~nm.
$\sopt$ was obtained from the third harmonic interferometric pseudo-heterodyne signal.\cite{Chen2012,Fei2012}
For simplicity most figures only show $\Re\sopt$, however similar information appears in $\Im\sopt$ as described by equation \eqref{eq:fit};
all analysis (background subtraction, fitting, etc.) was performed simultaneously on $\Re\sopt$ and $\Im\sopt$.
The location of the etched graphene edge ($x=0$) was determined from the simultaneously-measured topography.

The theoretical model of plasmon modes was calculated in a classical electromagnetic transfer matrix method,
with a thin film stack of vacuum--SiO$_2$(285~nm)--h-BN(46~nm)--graphene--h-BN(7~nm)--vacuum.
Thin film and nonlocal effects reduce $\Re\pq$ by $\sim 5$--20\% compared to infinite dielectric Drude model calculation (see Supplement).
The zero temperature random phase approximation (RPA) result\cite{Wunsch2006,Hwang2007,Principi2009a} was used for the graphene nonlocal conductivity $\sigma(k,\omega)$.
The permittivity model of Ref.~\onlinecite{Cai2007} was used for the h-BN films, modified to include dielectric losses based on Ref.~\onlinecite{Caldwell2014}.
The damping effect from dielectric losses shown in Fig~\ref{fig4} was also calculated in this method, taking phonon linewidths of 6.5~meV in-plane and 1.9~meV out-of-plane in the terminology of Ref.~\onlinecite{Caldwell2014}, and their origin is discussed further in the Supplement.
In Fig.~\ref{fig2}c and Fig.~\ref{fig2}d, the color quantity plotted is the imaginary part of the reflection coefficient of evanescent waves, evaluated at the top h-BN surface.
In these figures the damping has been modified (e.g., reduced dielectric loss) to enhance the visibility of modes---this does not significantly modify the mode locations.

}  

%

\section*{Acknowledgements}
{\small
It is a great pleasure to thank Joshua D. Caldwell, Javier García de Abajo, Andrea Tomadin and Leonid Levitov for many useful discussions.
This work used open source software (www.matplotlib.org, www.python.org).
F.H.L.K. acknowledges support by the Fundacio Cellex Barcelona, the ERC Career integration grant 294056 (GRANOP), the ERC starting grant 307806 (CarbonLight). 
F.H.L.K., M.P. and R.H. acknowledge support by the E.C. under Graphene Flagship (contract no. CNECT-ICT-604391).
A.P. and G.V. acknowledge DOE grant DE-FG02-05ER46203 and a Research Board Grant at the University of Missouri.
M.C. acknowledges the support of MIUR-FIRB2012 - Project HybridNanoDev (Grant No. RBFR1236VV).
M.P. acknowledges the Italian Ministry of Education, University, and Research (MIUR) through the programs “FIRB - Futuro in Ricerca 2010” - Project PLASMOGRAPH (Grant No. RBFR10M5BT) and “Progetti Premiali 2012” - Project ABNANOTECH. 
Y.G. and J.H. acknowledge support from the US Office of Naval Research N00014-13-1-0662.
}

\section*{Author contributions}
{\small
A.W. and M.B.L. performed the experiments, discussed the results and wrote the manuscript.
Y.G. fabricated the samples.
A.P. and M.C. provided the theory on different loss mechanisms.
P.A.-G. helped with measurements.
K.W. and T.T. synthesized the h-BN samples.
G.V., M.P., J.H., R.H. and F.H.L.K. supervised the work, discussed the results and co-wrote the manuscript.
All authors contributed to the scientific discussion and manuscript revisions.
}


\pagebreak
\newpage
\widetext
\cleardoublepage

\begin{center}
\textbf{\Large Supplementary Material: Highly confined low-loss plasmons in graphene--boron nitride heterostructures}
\end{center}
\setcounter{equation}{0}
\setcounter{figure}{0}
\setcounter{table}{0}
\setcounter{page}{1}
\makeatletter
\renewcommand{\theequation}{S\arabic{equation}}
\renewcommand{\thefigure}{S\arabic{figure}}
\renewcommand{\thetable}{S\arabic{table}}
\renewcommand{\thepage}{\roman{page}}
\renewcommand{\bibnumfmt}[1]{[S#1]}
\renewcommand{\citenumfont}[1]{S#1}

\section{Optical signal model}

Ideally, the tip interacts only with the local graphene underneath its apex, responding to the electric susceptibility of the graphene and acting as a localized ``point source'' for exciting the plasma wave.
The plasma wave spreads out as a circular wave (2D radial wave), reflects off the nearby edge of the graphene and returns to the tip.
Even in the ideal lossless case, only a small part of this returning wave couples to the tip, due to geometrical decay.
In practice, there are further interaction pathways: the light path does not only interact with the tip but also directly with the sample, and moreover the tip does not solely interact with the graphene under its apex.

This section describes the expected optical signal for a reflected circular wave, that has $\pwl/2$-period fringes as well as the origin of the fringes with $\pwl$-period and their expected optical signal.

\subsection{Signal in the bulk (local and launching response)}

In Fig.~2b of the main text, we demonstrate that the change in optical signal approximately follows the ac conductivity of the graphene. It is instructive to consider why this is, and why the correspondence might not be perfect.

To exactly calculate the optical signal measured in the s-SNOM is a complicated matter, however to first approximation, the incoming and outgoing light are only coupled to the charge oscillations in the metallized tip.
In the near-field limit, these charge oscillations are electrically (capacitively) coupled to the device under study.
In this picture, the optical signal is essentially related to part of the tip's self-capacitance that depends on tip-sample distance.

In the limit where the tip-sample system is non-resonant, the tip response can be calculated by some linear convolution of the surface's physical optical response.
In Fourier space:\cite{SFei2011}
\begin{equation}
 s(\omega) \approx \int w(k) r(\omega,k)\, dk,
\end{equation}
where $w(k)\sim k^2 \exp(-2kR)$ is a bell-shaped weighting function with a peak at $k \approx 10/\tipradius$, where $\tipradius$ is the tip radius.
The surface optical response is embedded in $r(\omega,k)$, the evanescent reflection coefficient for transverse magnetic waves having in-plane wavevector $k$ and angular frequency $\omega$.

In Sec.~\ref{sec:modesolver} we describe the general procedure to calculate $r(\omega,k)$ numerically, for an arbitrary stack.
In the quasi-electrostatic limit ($k \gg c/\omega$), we can write a simple expression for a dielectric-conductor-dielectric stack.
\begin{equation}
 r =
 \frac{\varepsilon-\varepsilon_0 - (\varepsilon+\varepsilon_0)\tfrac{\alpha}{1 - \alpha} e^{-2\eta k_x t}}
      {\varepsilon+\varepsilon_0 - (\varepsilon-\varepsilon_0)\tfrac{\alpha}{1 - \alpha} e^{-2\eta k_x t}},
 \label{eq:refl-BN-Gr-BN}
\end{equation}
Here, the upper dielectric is taken to have thickness $t$ and the lower to be infinite thickness, and the 2D conductor to be of zero thickness.
As the dielectric (h-BN) is anisotropic with in-plane permittivity $\varepsilon_{xx}$ much different from out-of-plane permittivity $\varepsilon_{zz}$, we have defined the effective permittivity,
\begin{equation}
 \varepsilon \equiv \sqrt{\varepsilon_{xx}\varepsilon_{zz}}
 ,
\label{eq:eff_perm}
\end{equation}
and the effective field confinement factor,
\begin{equation}
 \eta \equiv \sqrt{\frac{\varepsilon_{xx}}{\varepsilon_{zz}}}
 ,
\end{equation}
which in our experimental frequency range are $\varepsilon/\varepsilon_0 \approx  4.0\textrm{--}4.7$ and $\eta \approx 2$.
The effect of the graphene is captured in the parameter $\alpha$, defined as:
\begin{equation}
 \alpha \equiv \frac{\sigma}{2\varepsilon\omega i}k_x.
\label{eq:alpha-definition}
\end{equation}

\begin{figure}[t]
\centering
\includegraphics{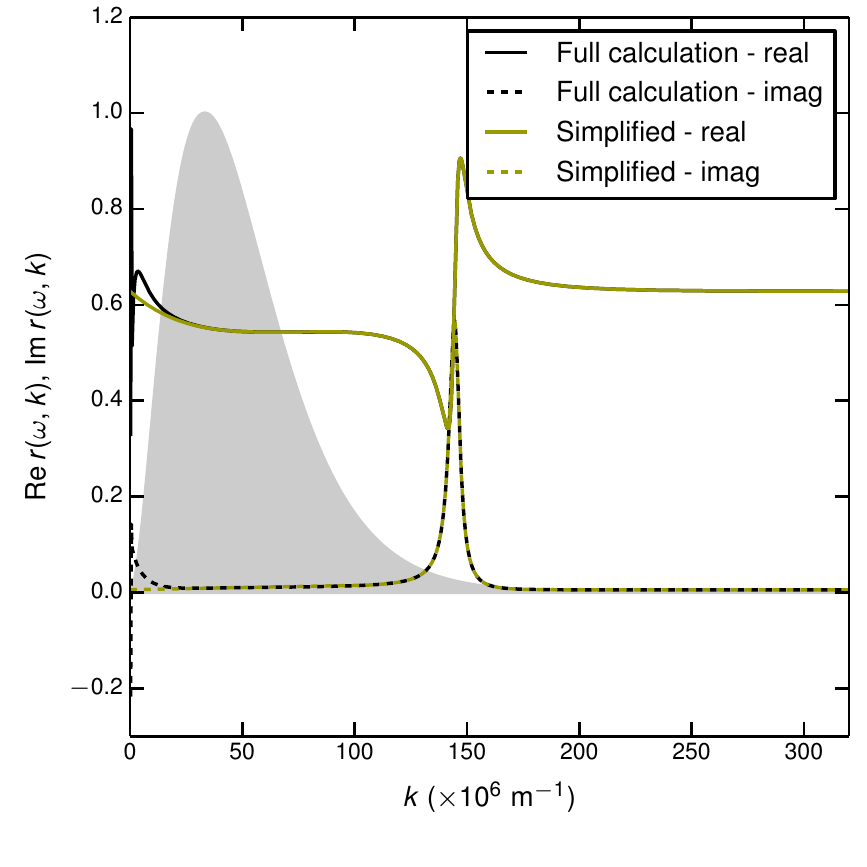}%
\includegraphics{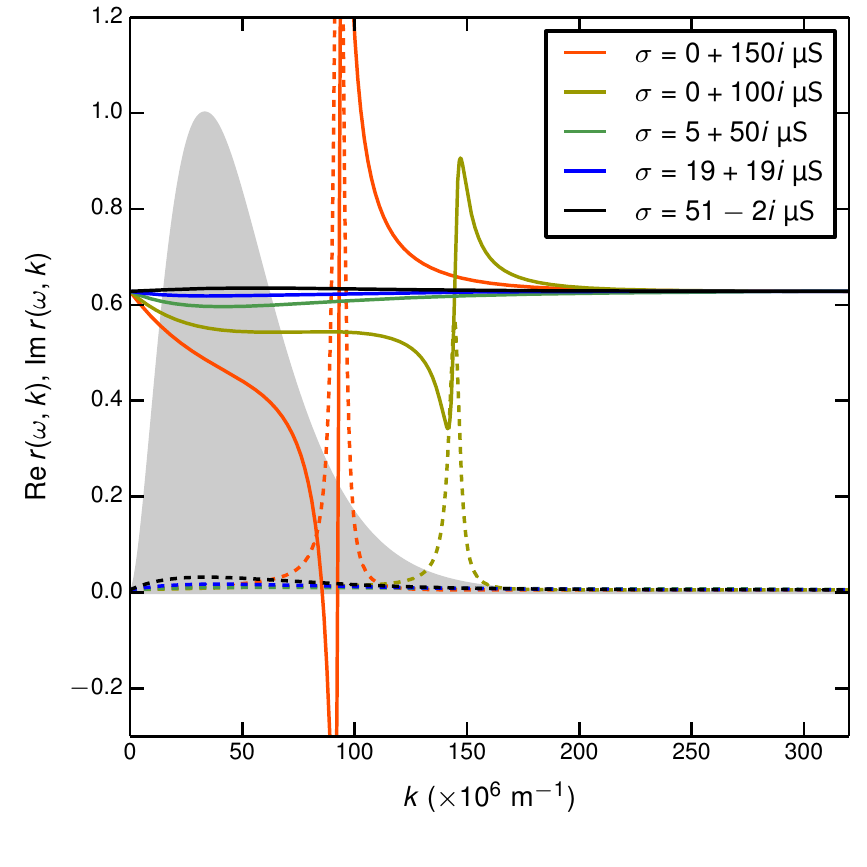}%
\caption{
\label{fig:reflection-coeff}
Dependence of the reflection coefficient on wave vector $k$.
a)
Comparison of the simple quasi-electrostatic result \eqref{eq:refl-BN-Gr-BN} for the BN-Gr-BN system, against the full electromagnetic calculation (Sec.~\ref{sec:modesolver}) that also includes the underlying SiO2.
The real part is shown as a solid line and the imaginary part as a dashed line.
Here the frequency is a typical $\frac{\omega}{2\pi} =30~\mathrm{THz}$, and for simplicity we have taken the conductivity of the graphene to be $\sigma = 10^{-4}i$~S, which corresponds to a carrier density of $n_\mathrm{s} \approx 2.6\times 10^{16}~\mathrm{m^{-2}}$
The filled-in curve shows the weighting function $w(x)$.
b)
Influence of varying the conductivity on \eqref{eq:refl-BN-Gr-BN}.
The conductivity variation here ranges from a higher carrier density ($n_\mathrm{s} \approx 5.0\times 10^{16}~\mathrm{m^{-2}}$ for orange curve) to zero carrier density (black curve).
For small $k_x$ values it is apparent that the shift in $r$ is proportional to $-\sigma/i$.
}
\end{figure}

Fig.~\ref{fig:reflection-coeff} plots the reflection coefficient for a typical frequency in our experiment.
As $k$ increases, the reflection coefficient probes the optical response closer and closer to the surface.
At very high $k \gtrsim 1/(\eta t) $ we only see the response of the top dielectric, which has the form:
\begin{equation}
r(k \rightarrow \infty) = r_\infty = \frac{\varepsilon-\varepsilon_0}{\varepsilon+\varepsilon_0}.
\end{equation}
In our case $r_\infty \approx 0.6$.
As $k$ is lowered, we meet at some point the condition $\alpha \approx 1$, leading to a resonance due to the denominators $1-\alpha$ in Eq.~\eqref{eq:refl-BN-Gr-BN}.
This is essentially the location of the plasmon; in fact the precise condition is a pole in $r$, which occurs slightly away from $\alpha=1$ due to the finite $t$ effects.
At the plasmon resonance, $r$ shows a strong peak in its imaginary part, indicating energy transfer to the plasmon.
This high-$k$ limit and the plasmon resonance are however both weakly coupled to the tip, being in the tail of the function $w(k)$.

The dominant contribution in our case comes from small $ k \approx 30\times 10^6~\mathrm{m^{-1}}$, which is generally below the plasmon resonance.
Being below the plasmon resonance, we can make the approximation $\alpha \ll 1$.
Expanding \eqref{eq:refl-BN-Gr-BN} to first order in $\alpha$, we find:
\begin{equation}
 r(k \ll \pq) \approx r_\infty - (1 - r_\infty{}^2) e^{-2\eta k_x t} \alpha.
\end{equation}
From this expression it is apparent why changes in graphene conductivity appear proportionally in our optical signal measurements.
For fixed frequency, the permittivity parameters $\varepsilon$, $\eta$, $r_\infty$ are fixed.
In essence, the tip can only couple well to small $k_x$ ($\alpha \ll 1$), and so {\em regardless of further details of the tip coupling}, the presence of the graphene causes a small perturbation that is proportional to its local conductivity.

Beyond the simple argument presented above, a number of further influences should be considered, and so we do not expect exact correspondence.
First, it is only to first order in $\alpha$ that the signal should be proportional to conductivity.
Higher order terms certainly do contribute, e.g., our imaging of plasmons requires the plasmon pole to contribute to the optical signal.
The plasmon draws energy from the tip and carries it away, and this energy loss appears similarly to dissipation (i.e., like $\Re\sigma$ or $\Im\varepsilon$).
As carrier density increases and the plasmon couples more efficiently, this energy loss becomes stronger.
Second, graphene can screen the influence of the dielectric layers underneath, in particular the SiO$_2$.
This screening effect also changes with carrier density, and so the influence of the SiO$_2$ is variable.

\subsection{Edge-reflected fringes ($\pwl/2$-period contribution)}

It is well known that in the far field, the amplitude of a lossless circular wave decays as $\sim 1/\sqrt{r}$, where $r$ is distance from source.
This ensures energy conservation on the wavefront, which has circumference $2\pi r$.
Mathematically, this appears in the 2D Helmholtz equation with point source at $\vec r_\mathrm{s}$,
$$ \nabla^2 E(\vec r) + q^2 E(\vec r) = -\delta(\vec r-\vec r_\mathrm{s}) $$
which has the solution
\begin{equation}
E(\vec r) = \tfrac{i}{4} H_0^{(1)} (q |\vec r-\vec r_\mathrm{s}|) 
\label{eq:green-hankel}
\end{equation}
where $H_0^{(1)}(z)$ is the first Hankel function of order zero.
In the case of plasmons, the wave field $E$ may represent charge density or out-of-plane electric field.
Equation \eqref{eq:green-hankel} remains a solution also when $q$ is complex, and describes a decaying wave for $\operatorname{Im}(q) > 0$.
As expected, the asymptotic decay of the Hankel function is  $H_0^{(1)}(z) \approx \sqrt{\frac{2}{i\pi z}} e^{i z}$.

Now, consider the case where the tip is near a straight edge -- the circular wave will reflect off this edge.
Assuming that the reflection coefficient is independent of the wave angle, then the reflected wave can be described using the mirror-image method.
Let the straight edge be defined by the line $r_x = 0$, and let the tip be at location $\vec r_\mathrm{tip} = (x,0)$.
Its mirror image is at $(-x,0) = -\vec r_\mathrm{tip}$.
The resulting total wave will be:
$$
E(\vec r) = E_{\rm launch} \tfrac{i}{4} H_0^{(1)} (q |\vec r - \vec r_\mathrm{tip}|) + E_{\rm refl} \tfrac{i}{4} H_0^{(1)} (q |\vec r + \vec r_\mathrm{tip}|) 
$$
The reflection coefficient $E_{\rm refl}/E_{\rm launch}$ is not necessarily unity. It is expected to be phase shifted\cite{SNikitin2014b} and also its magnitude will be smaller than unity due to energy loss from light emission and scattering at the edge.

We have used the s-SNOM in interferometric mode and so the measured signal is proportional to this complex field.\cite{SOcelic2006}
Ideally, the out-scattered light depends only on the local coupling to $E(\vec r_\mathrm{tip})$, and so $s \propto E(\vec r_\mathrm{tip})$.
The field from the first term (launched wave) forms part of the bulk signal.
The second term adds to the bulk signal and generates the interference fringes.
We thus expect:
$$ \xi(x) = \sbulk + A H_0^{(1)} (2 q x), $$
where $\sbulk$ collects together all contributions that would already occur away from the edge -- local response, plasmon launching, etc.,
and, the complex coefficient $A$ collects together factors of reflection, in-coupling, out-coupling, etc.

\subsection{Edge-launched fringes ($\pwl$-period contribution)} 

Broken translational symmetry at the edge provides for matching the small photon wavevector with the large plasmon wavevector.
As a simple model, one can think of the wave $E(x)$ being launched by an oscillating electric field at the edge.\cite{SZhang2014a}
This produces a plane wave plasmon without additional geometrical decay:
\begin{equation}
E(\vec r) \sim E_\mathrm{edge} e^{i \pq r_x},
\label{eq:edgelaunching-field}
\end{equation}
so that a contribution proportional to $E_\mathrm{edge} e^{i \pq x}$ is added to $\sopt(x)$.
This is the case for plasmons being launched directly by the illuminating laser spot which is effectively a plane wave on these nanometer length scales.

There are however other possibilities that lead to plasmons that travel only once the tip-edge distance $x$.
One possibility is the reverse of the above, that the plasmons launched at the tip are scattered to light at the graphene edge.
In this process the geometrical decay is less obvious: the plasma wave decays geometrically from the tip so that the field at the edge decays as $1/\sqrt{x}$, yet also the wave arrives in-phase over a larger section of the edge, tending to cancel this decay.

Another possibility is that the near-field tail of the tip interacts with the edge and launches a plasmon there, a plasmon which is then received at the tip after travelling $x$.
This is similar to the far-field case, except the tip acts as a field-enhancing mediator between light and edge.
Here additional geometrical decay is expected because the electric field of the near-field tail depends on the tip-edge distance.
It is not clear what distance dependence this near-field profile should take---monopolar, dipolar, or somewhere in-between.
This profile would also be modified by lateral field focussing by the h-BN.
Again, the reverse process (launching at tip, then the long-ranged tail of the edge plasmon field interacts with the tip) is also possible.

To allow for these various mechanisms we include a variable geometrical decay in this contribution to $\sopt$:
$$ \sedge(x) \propto \frac{e^{i \pq x}}{x^a + \tipradius^a},  $$
where $\tipradius$ is the tip apex radius, included to limit the divergence in this expression.
In the picture of plasmon plane wave launching at the edge, this would correspond to taking a distance-dependent edge field in Eq.~\eqref{eq:edgelaunching-field}
$$ E_\mathrm{edge}(x) \propto \frac{1}{x^a + \tipradius^a}.$$
The optical signal then shows the period of an edge-launched plane wave, but with additional geometrical decay whose origin is unclear.

\section{Fringe fitting (parameter extraction)}

In order to extract parameters, such as propagation length, from the fringe signal, we need an accurate model of the expected signal for a given amount of damping.
Based on the previous section, we have a decent model for the decay of fringes away from the edge:
\begin{equation}
\sopt(x) = \sbulk(x) +  A H_0^{(1)}(2 \pq x) + B \frac{e^{i \pq x}}{x^a + \tipradius^a}
\label{eq:fit-exact}
\end{equation}
where the fitting parameters are complex $A$, $B$ and $\pq$, and real $a$.
The tip radius is fixed to $\tipradius = 25$~nm.

There are some complications that prevent us from direct fitting of the raw data:
\begin{itemize}
 \item
 The location of the edge, $x=0$, needs to be detected in some way.
 \item
 The background part of the signal, $\sbulk(x)$, is not known a priori and we see clear signs of spatial variations.
 Fortunately, these variations (due to carrier density gradients) appear to be gradual.
 \item
 The model in Eq.~\eqref{eq:fit-exact} does not necessarily hold for small values of $x$.
 For the first fringe, the tip coupling mechanism may become very different than when the tip is over the bulk.
 Direct fitting of the data with equal residuals weighting is not suitable in this case.
\end{itemize}
The edge we detect from the topographic data of the s-SNOM apparatus, taking into account tip convolution effects.
To avoid biases from the unknown $\sbulk$ and the unknown first-fringe behaviour, we subtract a smooth background from the signal/model, and then perform fits in a transformed version of the signal/model.
In the following we describe this procedure in great detail.

\subsection{Detection of graphene edge location}

\begin{figure}[t]
\includegraphics{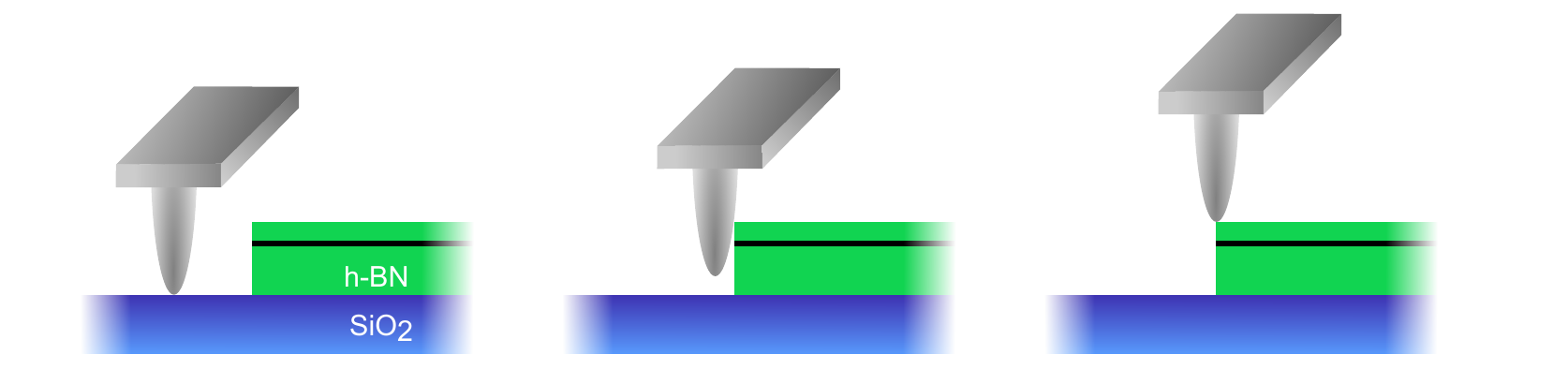}
\caption{\label{fig:supp-edge-detection}
Tip convolution effects make the topography edge appear away from the graphene edge. We define $x=0$ to occur when the tip is centered directly above the graphene edge---this is the situation depicted in the third panel.
\label{fig:edge-detection}
}
\end{figure}

\begin{figure}[t]
\centering
\includegraphics{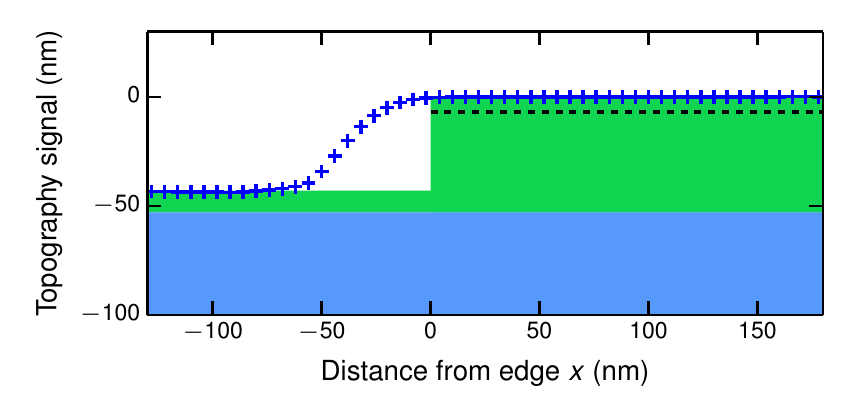}%
\caption{
\label{fig:edge-detection-topography}
Topographic signal from measurement (blue crosses) with the interpreted actual topography underneath.
The graphene (dashed line) is assumed to terminate at the end of the round feature, since this round feature is interpreted as a tip convolution effect.
}
\end{figure}

In separating out the contributions from geometrical decay from exponential decay, it is important that the location of $x=0$ (the graphene edge) has been determined with accuracy. 
An error in this determination leads to error in the extracted damping.

We have chosen to use our topographic data to determine this edge location.
It is well known that tip convolution artifacts result in modified appearances of sharp edges in scanning probe microscopy.
We assume that our physical etched edge is sharply vertical as illustrated in Fig.~\ref{fig:edge-detection}, such that the rounding and sloping apparent in the topographic signal is purely due to the AFM tip convolution (Fig.~\ref{fig:supp-edge-detection}).
As a result, the edge is located directly beneath the point where the rounding convolution ends, illustrated in Fig.~\ref{fig:edge-detection-topography}.
With the chosen edge-detection algorithm it is only possible that the graphene edge is actually further on the left in Fig.~\ref{fig:edge-detection-topography} which would lead to an underestimation of our extracted inverse damping ratios.
Note that even if the edge were not strictly vertical, the error in $x$ would be on the order of a few nanometers since the graphene lies only 7~nm under the surface.

\subsection{Background subtraction}

Since $\sbulk(x)$ is not known a priori, we can only estimate it from the dataset itself.
After discarding the data for $x < 0$, we estimate $\sbulk(x)$ by smoothing the measured $\sopt(x)$.
The difference,
\begin{equation}
\sdiff(x) = \sopt(x) - \ssmooth(x)
\end{equation}
should then be free of influence from the unknown $\sbulk$.

Background subtraction always results in removal of some of the desired signal, and is a well known source of statistical bias.
In this case, background subtraction leaves transient artifacts near $x=0$ due to the abrupt termination of the signal, and also selectively removes part of the fringes depending on their period (i.e., affecting more the $\lambda_\mathrm{p}$-period fringes than $\lambda_\mathrm{p}/2$-period fringes).
In order to give a fair comparison, we apply the same background subtraction procedure to the models used in the fit.

\subsection{Complex Hankel Transform}

Our goal is to access the asymptotic decay away from the edge, where we suppose tip coupling details are captured in the position-independent parameters $A$, $B$, $a$.
Close to the edge, such as with the first fringe, the tip coupling details are not necessarily this simple.
To this end, we perform fitting not in flat $\sdiff(x)$ space but rather in a transformed $\sdiff(x)$, in effect de-emphasizing the weight of the signal near the edge.
One possible approach here would be to take a Hankel transform of the $\sdiff(x)$; the Hankel transform is analogous to a Fourier transform, but more appropriate for circular symmetry, and it naturally gives stronger weight to a larger distance from origin.
The Hankel transform is however only a real transform, and does not mix together the real and imaginary parts of $\sdiff(x)$ in a convenient way.

We therefore use a ``complex Hankel transform'' of the following form:\cite{SCraeye1999}
\begin{equation} 
T(k) = \frac12 \int_0^\infty x [H_0^{(1)}(kx)]^* \sdiff(x) u(x) \, dx.
\label{eq:cht}
\end{equation}
This transform has the desirable property that $e^{iqx}$-type wave will transform to a peak near $+q$, and a $e^{-iqx}$ wave will transform to a peak near $-q$.
Note that unlike the proper Hankel transform, this transform is not simply invertible,\cite{SWengrovitz1987} however it is linear and successfully distinguishes $+q$ and $-q$ waves.
The function $u(x)$ in Eq.~\eqref{eq:cht} is a ``window'' function, used to select an appropriate range including sufficient fringes but without too much influence from noise.
We use the window $u(x) = 1 - \sin^2(\tfrac{\pi}{2}x/L)$ which produces a smooth cutoff as $x$ approaches $L$, with $L = \SI{1}{\micro\meter}$.

\begin{figure}[t]
\centering
\includegraphics{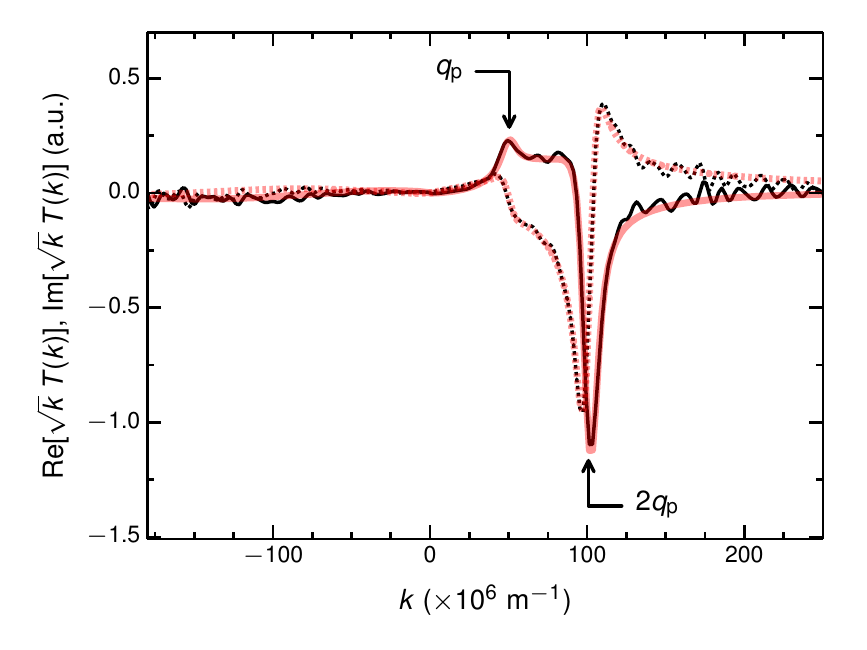}%
\caption{
\label{fig:hankel-fit}
Complex hankel transform the data in Fig.~3(a) in the main text. 
The real part are the solid curves and the imaginary part the dotted curves.
Black shows the measured data and red the fit.
}
\end{figure}

After applying Eq.~\eqref{eq:cht} to our background-subtracted and windowed fringes, we observe a function with two strong peaks (Fig.~\ref{fig:hankel-fit}), one peak at $\pq$ corresponding to processes where the plasmon travels $x$, and the other peak at $2\pq$ when the plasmon travels $2x$.
Note the absence of peaks at negative $k$, which confirms that we retrieve the phase of the light signal correctly and that the plasmons have positive group velocity.
The peak widths in Fig.~\ref{fig:hankel-fit} are related to the decay, though also affected by our choice of background subtraction and windowing procedures.
To obtain a fair comparison we can calibrate these peaks against a model with known decay.
What we do is to perform the same background subtraction, same windowing, and same transform on the model described in Eq.~\eqref{eq:fit-exact}.
We then fit the transformed model onto the transformed data, with equal weighting of the residuals $\sqrt{k}(T_\mathrm{data}(k)-T_\mathrm{model}(k))$ for equally-spaced $k$ values, over a specified $k$ range around the peaks.

While the above procedure may seem to overcomplicate matters, we stress that we have only performed a linear transformation on the data and model, and so we are effectively performing non-linear least squares on the source data but with modified residual weights.
The ultimate proof of this technique is the quality of fits (e.g., Fig.~3 of the main text, which is very good for the range of parameters presented in the manuscript.).
Besides being a reliable way to extract damping information, this technique also allows us to measure accurately the wavelength of fringes that are nearly invisible in the raw data.
An additional benefit is that we can directly visualize (Fig.~\ref{fig:hankel-fit}) that there are not additional components in the data as might correspond to $3\pq$, $-\pq$, etc., thereby confirming that the interferometric detection technique has precisely measured the light phasor.

\section{Mode calculations\label{sec:modesolver}}

We use the AC Maxwell equation for fields oscillating as $\exp(-i\omega t)$ in time,
\begin{equation}
 \vec\nabla \times (\vec\nabla \times \vec E) = \frac{\omega^2}{c^2} ( \vec E + \tfrac{1}{-i\omega\varepsilon_0} \vec J ),
\label{eq:maxwell-cw}
\end{equation}
with current given by
\begin{equation*}
 \vec J_\mathrm{diel} = -i \omega (\varepsilon - \varepsilon_0) \vec E
\end{equation*}
in the dielectrics (note $\varepsilon$ is a rank-2 tensor in h-BN), and by the nonlocal 2D conductivity relation
\begin{equation*}
 \vec J_\mathrm{gr}(\omega,x,y,z) = \delta(z-z_\mathrm{gr}) \iint dx'\,dy'\, \vec E(\omega,x',y',z_\mathrm{gr}) \sigma_\mathrm{NL}(\omega,x-x',y-y') 
\end{equation*}
in the graphene, where $z_\mathrm{gr}$ is the height of the graphene and $\sigma_\mathrm{NL}(\omega, x, y)$ is its nonlocal 2D conductivity function.

We neglect magnetic susceptibilities, whose bound currents would take the form
$
 \vec J_\mathrm{mag} = \frac{1}{i\omega} \vec\nabla \times [ (\mu_0^{-1} - \mu^{-1}) \vec\nabla \times \vec E],
$
i.e., we take the materials to be non-magnetic with permeability $\mu = \mu_0$.
In fact, even if the materials were slightly magnetic this would not influence the quasi-electrostatic limit described below.

We consider solutions that are plane waves along $x$ and constant along $y$, i.e., varying as $\exp(i k_x x + 0 y)$.
This reduces the system to a one dimensional problem in $z$, which we solve using the transfer matrix method.
There are two possible polarizations here: transverse magnetic ($E_y = 0$, $B_x = 0$, $B_z = 0$) and transverse electric ($B_y = 0$, $E_x = 0$, $E_z = 0$).
The tip couples essentially only to the transverse magnetic polarization, and plasmons only appear in this polarization.
We define the reflection coefficient $r(\omega, k_x)$ for transverse magnetic waves as the ratio of $E_z$ components of the up-decaying wave (positive $\Im k_z$) to the down-decaying wave (negative $\Im k_z$) at the top surface, with the condition that the wave is purely down-decaying at the bottom surface.

Bound propagating modes, such as the plasmon, appear in $r(\omega, k_x)$ as a simple pole in the complex $k_x$ plane, with a residue that is primarily real-valued.
Considering damped modes with $\Re k_x > 0$, then for ordinary dispersion (positive group velocity) this pole appears with $\Im k_x > 0$ and positive residue; for anomalous dispersion (negative group velocity), the pole has $\Im k_x < 0$ and negative residue.

For obtaining compact analytic equations such as \eqref{eq:refl-BN-Gr-BN} it is helpful to take the quasi-electrostatic approximation, where the effects of electromagnetic induction are neglected.
First we observe the relation $\vec\nabla\cdot ( \vec E + \tfrac{1}{-i\omega\varepsilon_0} \vec J ) = 0$, a consequence of taking the divergence of both sides of \eqref{eq:maxwell-cw}.
We then take the limit $c \rightarrow \infty$ (that is, $\mu, \mu_0 \rightarrow 0$, keeping $\varepsilon$ intact) which implies $\vec\nabla \times (\vec\nabla \times \vec E) = 0$.
This approximation is highly accurate when examining the near field waves (at very high $k_x$ values past the light line, i.e., where $k_x \gg 1/\sqrt{\varepsilon\mu}$).

\subsection{Nonlocal conductance}

The effects of 2D nonlocality are easy to include for plane waves, since in this case the convolution is converted into a $k_x$-dependent conductivity, $\sigma(\omega,k_x) = \iint dx\,dy\, e^{ik_x x}\sigma_\mathrm{NL}(\omega,x,y) $.
The quantity $\sigma(\omega,k_x)$ is known analytically at zero temperature, in the random phase approximation, allowing fast numerical evaluation.\cite{SKoppens_Nano_Lett_2011}
Although early works emphasized the influence of nonlocality,\cite{SWunsch2006,SHwang2007} we note a subtle point which is that for suspended graphene (in a vacuum dielectric) the primary nonlocal effect is the interband nonlocality, whereas for graphene in a dielectric and at low frequencies the dominant nonlocal effect is {\em intraband}.

We note that this intraband nonlocal effect is not particularly quantum nor special to graphene, but appears in all plasma physics.
For example, in the classical quasi-electrostatic plasma (Langmuir wave), microscopic thermal effects give a similar nonlocal conductivity.
For small $k$ the classical nonlocal conductivity takes the form
$$
\Im\sigma_{\rm classical}(\omega,k) \approx \omega^{-1} e^2 \frac{n}{m} [ 1 + 3 k^2 v_\mathrm{th}^2/\omega^2 ] ,
$$where $v_\mathrm{th} = \sqrt{kT/m}$ is the thermal speed.
The nonlocal effect can be seen as coming from pressurization effects, in the fluid plasma model.
From the point of view of kinetic theory (e.g., Vlasov equation) it is a consequence of near-resonant particles that are travelling close to the wave phase speed $\omega/k$, and is closely related to the Landau damping described below.
In a bulk plasma, the plasma condition is $\omega_\mathrm{p} = \sigma/(i \varepsilon)$, where $\sigma$ is the free-electron conductivity and $\varepsilon$ is the background permittivity from vacuum and bound electrons ($\varepsilon = \varepsilon_0$ in a gas plasma).
The nonlocal energy transport increases the imaginary part of conductivity and therefore increases the frequency of the plasma, as seen in the resulting Bohm-Gross dispersion,
$ \omega_\mathrm{p}^2 \approx e^2 \frac{n}{m \varepsilon_0} + 3 k^2 v_\mathrm{th}^2 $.

For the degenerate graphene electron gas we have for small $k$ and for $\omega \ll k_\mathrm{F} v_\mathrm{F}$,
$$
\Im\sigma_{\rm graphene}(\omega,k) \approx \omega^{-1}e^2 (2 v_\mathrm{F} k_\mathrm{F}/\hbar) [ 1 + \tfrac{3}{4} k^2 v_\mathrm{F}^2/\omega^2 ],
$$
where $k_\mathrm{F} = \sqrt{\pi n_\mathrm{s}}$ is the Fermi wavevector.
Again, this nonlocality can be interpreted as a near-resonant effect of electrons whose speed ($v_\mathrm{F}$) and direction are close to the wave phase speed $\omega/k$.
In a diagrammatic perturbation theory picture this corresponds to virtual intraband excitations.
The nonlocality occurs regardless of whether $k$ is comparable to $k_\mathrm{F}$.
The corresponding 2D plasma condition is $\omega_\mathrm{p} = \tfrac 12 \pq \sigma/(i \varepsilon)$, and so here too the nonlocal energy transport increases the imaginary part of conductivity and therefore increases the frequency of the plasma, or for fixed frequency it lowers $\pq$.
In fact the full expression for $\sigma_{\rm graphene}$ contains a diverging conductivity as $k$ approaches $\omega/v_\mathrm{F}$.
This prevents the plasmon from having a lower phase velocity than $v_\mathrm{F}$. 

The striking difference between the classical thermal plasma and the graphene plasma is the effect of Landau damping.
In the thermal plasma, the thermal distribution implies that some electrons have a velocity as high as the plasma phase velocity.
They `surf' the wave, accelerating to higher speeds and drawing energy out from the plasma.
Thus, nonlocal effects in a thermal plasma are rarely observed because Landau damping turns on at the same time.

For plasmas in degenerate electron systems such as metals, low-temperature doped semiconductors, or doped graphene, it is possible to see the nonlocality without Landau damping, since the Landau damping only turns on after the plasmon wavevector $\pq$ passes above $\omega/v_\mathrm{F}$.
This is because the electrons have a sharp cutoff in their speed distribution, with few electrons travelling faster than $v_\mathrm{F}$.
Still, the nonlocality can be difficult to access: in semiconductors for example the Fermi velocity is very low, and so the required $\pq$ to observe nonlocality are quite high.
In graphene, the high Fermi velocity allows easier observation of nonlocality in plasmonics.
This is most apparent with a high permittivity environment around the graphene, since for frequencies below $k_\mathrm{F} v_\mathrm{F}$ this drives the plasmon into the intraband nonlocal regime before it is affected by the interband absorption.

\subsection{Dispersion relation comparison}

\begin{figure}[t]
\centering
\subfigure{\label{fig:supp-fdep-k-drude}\includegraphics{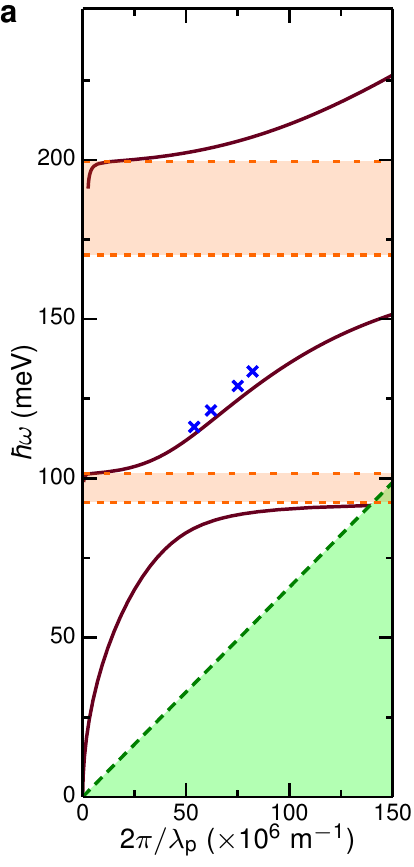}}%
\subfigure{\label{fig:supp-fdep-k-drude-thinfilm}\includegraphics{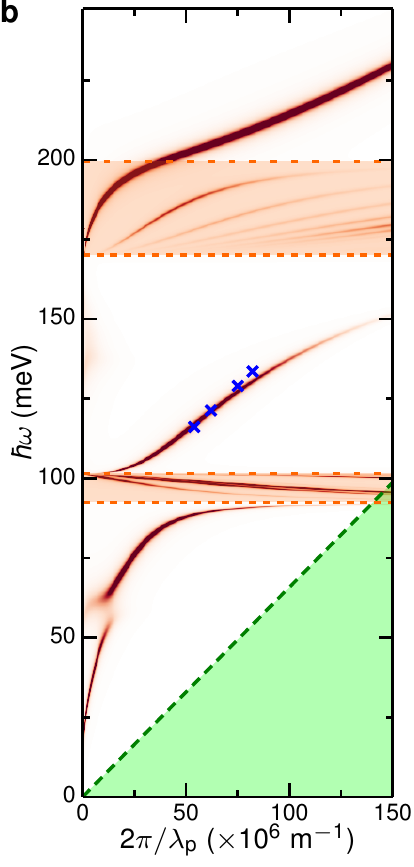}}%
\subfigure{\label{fig:supp-fdep-k-nonlocalrpa}\includegraphics{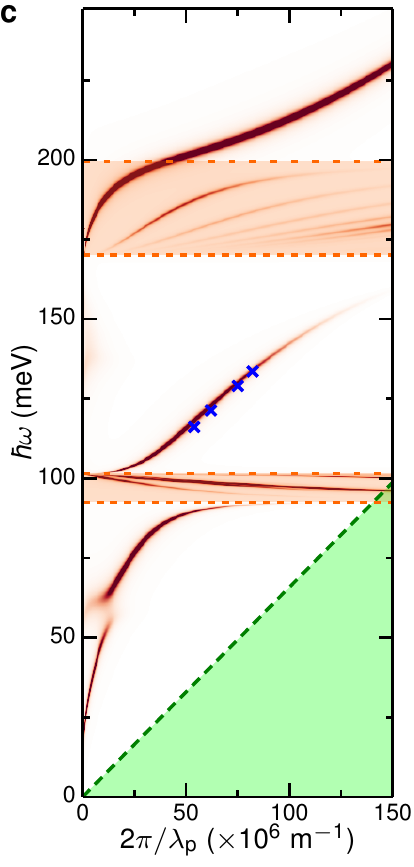}}%
\caption{
\label{fig:dispersion_comparison_f}
(a) Drude model for graphene conductivity and simple effective permittivity for h-BN surrounding the graphene. (b) Drude model for graphene conductivity and thin film effects for h-BN surrounding the graphene. (c) Non local RPA for graphene conductivity and thin film effects for h-BN surrounding the graphene. Scattering time $\tau = 500~\mathrm{fs}$, $n_s=7.37\times10^{12}~\mathrm{cm^{-2}}$.
}
\end{figure}

\begin{figure}[t]
\centering
\subfigure{\label{fig:supp-gatedep-k-drude}\includegraphics{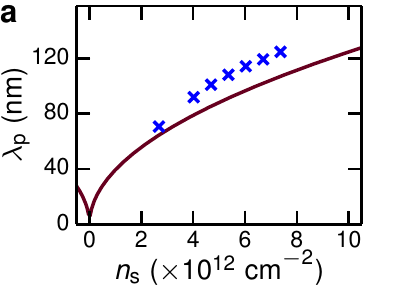}}%
\subfigure{\label{fig:supp-gatedep-k-drude-thinfilm}\includegraphics{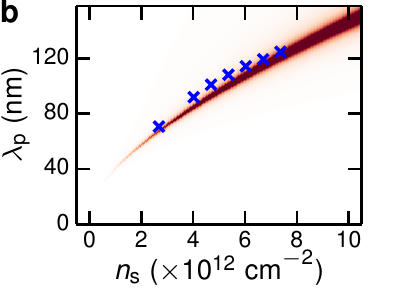}}%
\subfigure{\label{fig:supp-gatedep-k-nonlocalrpa}\includegraphics{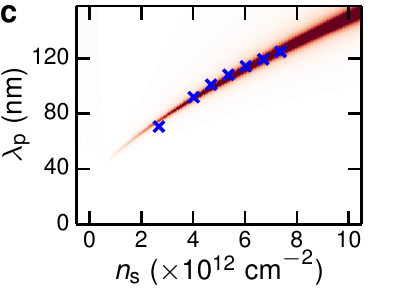}}%
\caption{
\label{fig:dispersion_comparison_n_s}
(a) Drude model for graphene conductivity and simple effective permittivity for h-BN surrounding the graphene. (b) Drude model for graphene conductivity and thin film effects for h-BN surrounding the graphene. (c) Non local RPA for graphene conductivity and thin film effects for h-BN surrounding the graphene. Scattering time $\tau = 500~\mathrm{fs}$, $n_s=7.37\times10^{12}~\mathrm{cm^{-2}}$.
}
\end{figure}

In the simple Drude model the local response conductivity is given by the following expression:\cite{SKoppens_Nano_Lett_2011}
\begin{equation}
\sigma(\omega,\tau,n_s)=\dfrac{2e^2v_\mathrm{F}}{h}\dfrac{\sqrt{\pi n_s}}{1/\tau-i\omega}
\label{eq:Drude}
\end{equation}

In Fig.~\ref{fig:dispersion_comparison_f} and Fig.~\ref{fig:dispersion_comparison_n_s} we compare different dispersion models.
In Fig.~\ref{fig:dispersion_comparison_f},\ref{fig:dispersion_comparison_n_s}a we show the result for a simple Drude conductivity for the graphene as in \eqref{eq:Drude} and the simple graphene plasmon relation\cite{SJablan_PRB_2009} $\pq \approx 2\omega\epsilon(\omega) i/\sigma(\omega)$, where $\omega$ is the angular frequency of the excitation light, $\epsilon$ is the effective permittivity of the dielectric environment -- see \eqref{eq:eff_perm} above -- and $\sigma$ is the local conductivity as defined in \eqref{eq:Drude}.
Both top and bottom h-BN layer are considered to be semi-infinite.
Note that here no propagating phonon polariton modes exist inside the reststrahlen bands marked in orange.\cite{SPrincipi}

A significantly improved fit is achieved by using the Drude model but including thin film effects of the 46~nm bottom and 7~nm top h-BN in Fig.~\ref{fig:dispersion_comparison_f},\ref{fig:dispersion_comparison_n_s}b.
Due to the thin film effects propagating phonon polariton modes exist in the reststrahlen bands.
These modes will be discussed elsewhere.

Including the nonlocal conductivity in Fig.~\ref{fig:dispersion_comparison_f},\ref{fig:dispersion_comparison_n_s}c we achieve an even better fit, especially for higher wavevectors where nonlocal effects start playing a more significant role.
Also in the carrier density dependence in Fig.~\ref{fig:dispersion_comparison_n_s}c a significantly improved fit is achieved as compared to not including nonlocal effects in Fig.~\ref{fig:dispersion_comparison_n_s}b.

\section{h-BN permittivity model}

Hexagonal boron nitride is an anisotropic material and so its permittivity $\varepsilon$ is a tensor.
Choosing $x,y$ to be the in-plane directions and $z$ to be the out-of-plane direction (``$c$-axis''), by symmetry the permittivity must be diagonal in a perfect h-BN crystal:
$$
\varepsilon =
\begin{pmatrix}
\varepsilon_x & 0 & 0 \\
0 & \varepsilon_y & 0 \\
0 & 0 & \varepsilon_z \\
\end{pmatrix}
$$
with components $\varepsilon_{x} = \varepsilon_{y} \neq \varepsilon_{z}$.

As with many dielectric materials, the permittivity of h-BN is frequency dependent with resonances due to internal polar degrees of freedom,
\begin{equation}
\varepsilon_l(\omega) = \varepsilon_l(\infty) + s_{{\rm v},l} \frac{ \omega_{{\rm v},l}^2 }{\omega_{{\rm v},l}^2 - i \gamma_{{\rm v},l}\omega - \omega^2}, \qquad l = x,y,z
.
\label{eq:permittivity-lorentzian}
\end{equation}
This degree of freedom is a polar lattice vibration, and its permittivity contribution involves real-valued constants $s_{{\rm v},l}$ (dimensionless coupling factor), $\omega_{{\rm v},l}$ (normal frequency of vibration), and $\gamma_{{\rm v},l}$ (amplitude decay rate).
Observe that $s_{{\rm v},l}$ gives the DC permittivity contribution of the polar lattice distortion, so that in the case of a single vibrational mode as in \eqref{eq:permittivity-lorentzian}, one has $s_{{\rm v},l} = \varepsilon_{{\rm v},l}(0) - \varepsilon_{{\rm v},l}(\infty)$.
We neglect nonlocal effects in the permittivity of h-BN as they should only appear once $k$ is comparable to the reciprocal lattice vectors, a regime that is two orders of magnitude away from the experimental case.

In h-BN, it is theoretically expected that there are only three polar vibrational modes, one each for $x,y,z$.\cite{SGeick1966a}
The out-of-plane vibration ($l = z$) has significantly different values of $s_{{\rm v},l}, \omega_{{\rm v},l}, \gamma_{{\rm v},l}$ compared to the in-plane modes ($l = x,y$).
In practice, it is sometimes useful to include additional modes to fit the measured permittivity in disordered crystals,\cite{SGeick1966a} however here we consider ideal h-BN with one mode along each direction.

The permittivity \eqref{eq:permittivity-lorentzian} completely characterizes the h-BN for optical studies at frequencies up to and including the mid-infrared.
For bulk h-BN this permittivity is known to produce interesting behaviour of electromagnetic modes since $\Re \varepsilon_z \leq 0$ for one frequency band, and $\Re \varepsilon_x, \Re \varepsilon_y \leq 0$ in another frequency band.
Both frequency bands contain:
\begin{itemize}
 \item Transverse phonon polaritons near $\omega_{{\rm v},l}$. Near this frequency, the permittivity along $l$ diverges to very large values ($\Re \varepsilon_l \sim \pm 100$). For light polarized along direction $l$, a strong peak in reflectivity (near 100\%) is observed at this frequency.\cite{SGeick1966a, SCaldwell2014}
 \item Longitudinal phonon polaritons near $\omega_{{\rm L},l} = \omega_{{\rm v},l} \sqrt{\varepsilon_l(0)/\varepsilon_l(\infty)}$. At this frequency, $\varepsilon_l$ passes close to 0. This allows a purely electric oscillation that is longitudinal, i.e., electric field parallel with the phase velocity.\cite{SGeick1966a}
 \item Hyperbolic phonon polaritons for $\omega_{{\rm v},l} < \omega < \omega_{{\rm L},l}$. In this frequency range, $\Re \varepsilon_l < 0$ in direction $l$, yet $\Re \varepsilon$ is positive along another direction. This results in a hyperboloidal constant-frequency surface of propagating modes in $k$-space, rather than the usual ellipsoid that appears for most frequencies. This hyperboloid extends to very high $k$ (short wavelength) allowing propagating modes of very short wavelength.\cite{SCaldwell2014} The group velocities of these confined modes are correspondingly low and are nearly perpendicular to their phase velocities.
\end{itemize}
These special frequency intervals are marked in Fig.~2 of the main text as orange bands:
In the lower frequency band, $\Re \varepsilon_z <0 $ whereas $\Re \varepsilon_{x,y} <0 $ in the higher frequency band.
For thin h-BN films, the effects of the transverse and longitudinal modes are somewhat diminished, yet the hyperbolic modes start to exhibit waveguiding\cite{SDai2014b} and have been exploited to produce subwavelength resonant structures.\cite{SCaldwell2014}

For the frequencies investigated in this study, $\Re \varepsilon$ is strictly positive and the most important aspect of \eqref{eq:permittivity-lorentzian} is its anisotropy and its dielectric loss.
The overall permittivity, including its high anisotropy (with $\varepsilon_x \approx 9$ and $\varepsilon_z \approx 2$ in the studied frequency range), is important for matching the measured plasmon wavelengths.
Understanding the dielectric loss is a crucial part of understanding our plasmon damping.
The following subsections describe the parameter sets we have considered and how dielectric loss may be modified in thin films of h-BN.

\subsection{Parameters of h-BN permittivity and remarks}

\begin{table}[t]
\begin{minipage}[t]{0.7\textwidth}
\begin{tabular*}{1.0\columnwidth}{@{\extracolsep{\fill} } l|lrrrrrr}
\hline
Model & $l$  & $\varepsilon_l(\infty)$
  & \multicolumn{1}{c}{$s_{{\rm v},l}$} & $\hbar\omega_{{\rm v},l}/$meV & $\hbar\gamma_{{\rm v},l}/$meV \\
  \hline

Geick\cite{SGeick1966a} & $x,y$ & 4.95
 & 1.868 & 169.5 & 3.6 \\
\multicolumn{3}{r}{+}
 & 0.209 & 95.1 & 3.4 \\
 & $z$ & 4.10
 & 0.530 & 97.1 & 1.0 \\
\multicolumn{3}{r}{+}
 & 0.456 & 187.2 & 9.9 \\
\hline

Cai\cite{SCai2007} & $x,y$ & 4.87
 & 1.83 & 170.1 & --- \\
 & $z$ & 2.95
 & 0.61 & 92.5 & --- \\
\hline


Caldwell\cite{SCaldwell2014} & $x,y$ & 4.9
 & 2.001 & 168.6 & 0.87 \\
 & $z$ & 2.95
 & 0.5262 & 94.2 & 0.25 \\
\hline

Cai "clean" & $x,y$ & 4.87
 & 1.83 & 170.1 & 0.87 \\
 & $z$ & 2.95
 & 0.61 & 92.5 & 0.25 \\
\hline

Cai "damaged" & $x,y$ & 4.87
 & 1.83 & 170.1 & 6.5 \\
 & $z$ & 2.95
 & 0.61 & 92.5 & 1.9 \\
\hline

\end{tabular*} 
\end{minipage}

\caption{\label{permtab}
Different permittivity models for hexagonal boron nitride.
Note that the model of Geick~et~al. includes two vibrational modes for each direction. }
\end{table}

For reference, we list five permittivity parameter sets in Table~\ref{permtab}.
The first three models are from the existing literature \cite{SGeick1966a, SCai2007, SCaldwell2014} and the last two are hybrids that we have constructed.
In future precision studies it may be necessary to take into account (or exploit) the isotope effect of boron which could allow tuning of the resonance frequencies by 3\%, or to remove the dielectric loss that originates from the isotope inhomogeneity of natural boron.

The Geick~et~al. study was performed on a large h-BN sample, and the authours found it necessary to include an additional vibrational mode for each direction, in order to fit their reflectance data.
They attributed this necessity to the large degree of axis misalignment among the crystallites, which would mix together the $x$ and $z$ permittivities.\cite{SGeick1966a}

The Cai~et~al. model is a theoretical calculation for perfect h-BN,\cite{SCai2007} and results a plasmon dispersion that matches closely to the experiment.
The values of Cai~et~al. were used successfuly in modelling the propagating phonon polaritons in Ref.~\onlinecite{SDai2014b} (see the supplement of that paper).
This study does not address the expected dielectric losses.

Caldwell~et~al. present their values in the supplementary material of Ref.~\onlinecite{SCaldwell2014}.
This permittivity was inferred from reflectance measurements on thin h-BN exfoliated films, originating from the same source as the h-BN films in our study.
The parameters obtained here were very similar to the Cai~et~al. values.

Cai ``clean'' and Cai ``damaged'' in Table~\ref{permtab} take the theoretical modes of Ref.~\onlinecite{SCai2007} and incorporates empirical losses based on Refs.~\onlinecite{SCaldwell2014,SGeick1966a}.
Cai ``clean'' uses the losses for pristine thick films of h-BN as measured in Ref.~\onlinecite{SCaldwell2014}.
In Cai ``damaged'' we amplify these losses to appear similar to those observed for thin ($<$200~nm) h-BN films in the same work.
As no data were available on the losses of the $z$ vibrational mode for thin films, we assume that they increase in proportion with the $x,y$ losses, as described in the next section.

The model Cai ``clean'' was used to produced dispersion plots where we have matched the measured plasmon wavelength; its low level of dielectric loss aids the visibility of the modes.
This last model, Cai ``damaged'', was used in our calculations of plasmon damping.

\subsection{h-BN losses in thin films}

\begin{figure}[t]
\centering
\includegraphics{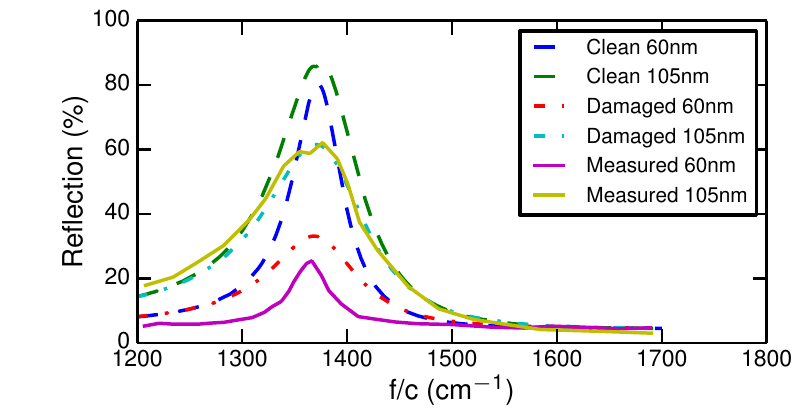}%
\caption{
\label{fig:BN-thin-film}
Comparison between measured reflection from the supplement of Ref.~\onlinecite{SCaldwell2014} and simulated reflection for thin h-BN flakes.
The clean h-BN uses an in-plane phonon linewidth of 0.87~meV from Ref.~\onlinecite{SCaldwell2014}.
The simulations of the damaged ones have an increased in-plane phonon linewidth of 3.7~meV in the case of the 105~nm thick h-BN and 6.5~meV for the 60~nm h-BN.
}
\end{figure}

Caldwell~et~al. have noticed in measuring thin h-BN films in Ref.~\onlinecite{SCaldwell2014} that the effective $\gamma_{{\rm v},l}$ seems to be larger than bulk. 
As a result our h-BN, especially the thin upper layer, may have higher losses. 
In Fig.~\ref{fig:BN-thin-film} a comparison between clean and damaged h-BN from Ref.~\onlinecite{SCaldwell2014} is made. 
The simulations were done using the transfer matrix method taking into account the thickness of the flakes and the substrate and for the BaF$_2$ substrate permittivity values from Ref.~\onlinecite{Palik1997} are used.
It is clear that the resonance linewidth reported for thicker ($>$200~nm) h-BN flakes becomes broadened for thinner flakes and strongly depends on the thickness.
Therefore the dielectric losses of graphene plasmons due to h-BN heavily depend on sample geometry and surrounding flake thickness.
The values used in Fig.~4 in the main text are shown as Cai "damaged" in Table~\ref{permtab}.
The out-of plane phonon width was estimated by using a ratio of 3.5 between in-plane and out-of-plane width as reported in Refs.~\onlinecite{SCaldwell2014} and~\onlinecite{SGeick1966a} for both very clean mono crystalline h-BN as well as for polycrystalline h-BN.
Considering the strong thickness dependence of the phonon linewidth as seen in Fig.~\ref{fig:BN-thin-film} these values are realistic.

%

\begin{thebibliography}{32}%
\makeatletter
\providecommand \@ifxundefined [1]{%
 \@ifx{#1\undefined}
}%
\providecommand \@ifnum [1]{%
 \ifnum #1\expandafter \@firstoftwo
 \else \expandafter \@secondoftwo
 \fi
}%
\providecommand \@ifx [1]{%
 \ifx #1\expandafter \@firstoftwo
 \else \expandafter \@secondoftwo
 \fi
}%
\providecommand \natexlab [1]{#1}%
\providecommand \enquote  [1]{``#1''}%
\providecommand \bibnamefont  [1]{#1}%
\providecommand \bibfnamefont [1]{#1}%
\providecommand \citenamefont [1]{#1}%
\providecommand \href@noop [0]{\@secondoftwo}%
\providecommand \href [0]{\begingroup \@sanitize@url \@href}%
\providecommand \@href[1]{\@@startlink{#1}\@@href}%
\providecommand \@@href[1]{\endgroup#1\@@endlink}%
\providecommand \@sanitize@url [0]{\catcode `\\12\catcode `\$12\catcode
  `\&12\catcode `\#12\catcode `\^12\catcode `\_12\catcode `\%12\relax}%
\providecommand \@@startlink[1]{}%
\providecommand \@@endlink[0]{}%
\providecommand \url  [0]{\begingroup\@sanitize@url \@url }%
\providecommand \@url [1]{\endgroup\@href {#1}{\urlprefix }}%
\providecommand \urlprefix  [0]{URL }%
\providecommand \Eprint [0]{\href }%
\providecommand \doibase [0]{http://dx.doi.org/}%
\providecommand \selectlanguage [0]{\@gobble}%
\providecommand \bibinfo  [0]{\@secondoftwo}%
\providecommand \bibfield  [0]{\@secondoftwo}%
\providecommand \translation [1]{[#1]}%
\providecommand \BibitemOpen [0]{}%
\providecommand \bibitemStop [0]{}%
\providecommand \bibitemNoStop [0]{.\EOS\space}%
\providecommand \EOS [0]{\spacefactor3000\relax}%
\providecommand \BibitemShut  [1]{\csname bibitem#1\endcsname}%
\let\auto@bib@innerbib\@empty
\bibitem [{\citenamefont {Fang}\ \emph {et~al.}(2014)\citenamefont {Fang},
  \citenamefont {Wang}, \citenamefont {Schlather}, \citenamefont {Liu},
  \citenamefont {Ajayan}, \citenamefont {{Garc\'{\i}a de Abajo}}, \citenamefont
  {Nordlander}, \citenamefont {Zhu},\ and\ \citenamefont {Halas}}]{Fang2014a}%
  \BibitemOpen
  \bibfield  {author} {\bibinfo {author} {\bibfnamefont {Z.}~\bibnamefont
  {Fang}}, \bibinfo {author} {\bibfnamefont {Y.}~\bibnamefont {Wang}}, \bibinfo
  {author} {\bibfnamefont {A.~E.}\ \bibnamefont {Schlather}}, \bibinfo {author}
  {\bibfnamefont {Z.}~\bibnamefont {Liu}}, \bibinfo {author} {\bibfnamefont
  {P.~M.}\ \bibnamefont {Ajayan}}, \bibinfo {author} {\bibfnamefont {F.~J.}\
  \bibnamefont {{Garc\'{\i}a de Abajo}}}, \bibinfo {author} {\bibfnamefont
  {P.}~\bibnamefont {Nordlander}}, \bibinfo {author} {\bibfnamefont
  {X.}~\bibnamefont {Zhu}}, \ and\ \bibinfo {author} {\bibfnamefont {N.~J.}\
  \bibnamefont {Halas}},\ }\href {\doibase 10.1021/nl404042h} {\bibfield
  {journal} {\bibinfo  {journal} {Nano Lett.}\ }\textbf {\bibinfo {volume}
  {14}},\ \bibinfo {pages} {299} (\bibinfo {year} {2014})}\BibitemShut
  {NoStop}%
\bibitem [{\citenamefont {Grigorenko}\ \emph {et~al.}(2012)\citenamefont
  {Grigorenko}, \citenamefont {Polini},\ and\ \citenamefont
  {Novoselov}}]{Grigorenko2012}%
  \BibitemOpen
  \bibfield  {author} {\bibinfo {author} {\bibfnamefont {A.~N.}\ \bibnamefont
  {Grigorenko}}, \bibinfo {author} {\bibfnamefont {M.}~\bibnamefont {Polini}},
  \ and\ \bibinfo {author} {\bibfnamefont {K.~S.}\ \bibnamefont {Novoselov}},\
  }\href {\doibase 10.1038/nphoton.2012.262} {\bibfield  {journal} {\bibinfo
  {journal} {Nature Photon.}\ }\textbf {\bibinfo {volume} {6}},\ \bibinfo
  {pages} {749} (\bibinfo {year} {2012})}\BibitemShut {NoStop}%
\bibitem [{\citenamefont {Gullans}\ \emph {et~al.}(2013)\citenamefont
  {Gullans}, \citenamefont {Chang}, \citenamefont {Koppens}, \citenamefont
  {{Garc\'{\i}a de Abajo}},\ and\ \citenamefont {Lukin}}]{Gullans2013}%
  \BibitemOpen
  \bibfield  {author} {\bibinfo {author} {\bibfnamefont {M.}~\bibnamefont
  {Gullans}}, \bibinfo {author} {\bibfnamefont {D.}~\bibnamefont {Chang}},
  \bibinfo {author} {\bibfnamefont {F.~H.~L.}\ \bibnamefont {Koppens}},
  \bibinfo {author} {\bibfnamefont {F.~J.}\ \bibnamefont {{Garc\'{\i}a de
  Abajo}}}, \ and\ \bibinfo {author} {\bibfnamefont {M.}~\bibnamefont
  {Lukin}},\ }\href {\doibase 10.1103/PhysRevLett.111.247401} {\bibfield
  {journal} {\bibinfo  {journal} {Phys. Rev. Lett.}\ }\textbf {\bibinfo
  {volume} {111}},\ \bibinfo {pages} {247401} (\bibinfo {year}
  {2013})}\BibitemShut {NoStop}%
\bibitem [{\citenamefont {Koppens}\ \emph {et~al.}(2011)\citenamefont
  {Koppens}, \citenamefont {Chang},\ and\ \citenamefont {{Garc\'{\i}a de
  Abajo}}}]{Koppens_Nano_Lett_2011}%
  \BibitemOpen
  \bibfield  {author} {\bibinfo {author} {\bibfnamefont {F.~H.~L.}\
  \bibnamefont {Koppens}}, \bibinfo {author} {\bibfnamefont {D.~E.}\
  \bibnamefont {Chang}}, \ and\ \bibinfo {author} {\bibfnamefont {F.~J.}\
  \bibnamefont {{Garc\'{\i}a de Abajo}}},\ }\href {\doibase 10.1021/nl201771h}
  {\bibfield  {journal} {\bibinfo  {journal} {Nano Lett.}\ }\textbf {\bibinfo
  {volume} {11}},\ \bibinfo {pages} {3370} (\bibinfo {year}
  {2011})}\BibitemShut {NoStop}%
\bibitem [{\citenamefont {Fei}\ \emph {et~al.}(2012)\citenamefont {Fei},
  \citenamefont {Rodin}, \citenamefont {Andreev}, \citenamefont {Bao},
  \citenamefont {McLeod}, \citenamefont {Wagner}, \citenamefont {Zhang},
  \citenamefont {Zhao}, \citenamefont {Thiemens}, \citenamefont {Dominguez},
  \citenamefont {Fogler}, \citenamefont {{Castro Neto}}, \citenamefont {Lau},
  \citenamefont {Keilmann}, \citenamefont {Basov},\ and\ \citenamefont
  {Castro-Neto}}]{Fei2012}%
  \BibitemOpen
  \bibfield  {author} {\bibinfo {author} {\bibfnamefont {Z.}~\bibnamefont
  {Fei}}, \bibinfo {author} {\bibfnamefont {A.~S.}\ \bibnamefont {Rodin}},
  \bibinfo {author} {\bibfnamefont {G.~O.}\ \bibnamefont {Andreev}}, \bibinfo
  {author} {\bibfnamefont {W.}~\bibnamefont {Bao}}, \bibinfo {author}
  {\bibfnamefont {A.~S.}\ \bibnamefont {McLeod}}, \bibinfo {author}
  {\bibfnamefont {M.}~\bibnamefont {Wagner}}, \bibinfo {author} {\bibfnamefont
  {L.~M.}\ \bibnamefont {Zhang}}, \bibinfo {author} {\bibfnamefont
  {Z.}~\bibnamefont {Zhao}}, \bibinfo {author} {\bibfnamefont {M.}~\bibnamefont
  {Thiemens}}, \bibinfo {author} {\bibfnamefont {G.}~\bibnamefont {Dominguez}},
  \bibinfo {author} {\bibfnamefont {M.~M.}\ \bibnamefont {Fogler}}, \bibinfo
  {author} {\bibfnamefont {A.~H.}\ \bibnamefont {{Castro Neto}}}, \bibinfo
  {author} {\bibfnamefont {C.~N.}\ \bibnamefont {Lau}}, \bibinfo {author}
  {\bibfnamefont {F.}~\bibnamefont {Keilmann}}, \bibinfo {author}
  {\bibfnamefont {D.~N.}\ \bibnamefont {Basov}}, \ and\ \bibinfo {author}
  {\bibfnamefont {A.~H.}\ \bibnamefont {Castro-Neto}},\ }\href {\doibase
  10.1038/nature11253} {\bibfield  {journal} {\bibinfo  {journal} {Nature}\
  }\textbf {\bibinfo {volume} {487}},\ \bibinfo {pages} {82} (\bibinfo {year}
  {2012})}\BibitemShut {NoStop}%
\bibitem [{\citenamefont {Chen}\ \emph {et~al.}(2012)\citenamefont {Chen},
  \citenamefont {Badioli}, \citenamefont {Alonso-Gonz\'{a}lez}, \citenamefont
  {Thongrattanasiri}, \citenamefont {Huth}, \citenamefont {Osmond},
  \citenamefont {Spasenovi\'{c}}, \citenamefont {Centeno}, \citenamefont
  {Pesquera}, \citenamefont {Godignon}, \citenamefont {{Zurutuza Elorza}},
  \citenamefont {Camara}, \citenamefont {{Garc\'{\i}a de Abajo}}, \citenamefont
  {Hillenbrand},\ and\ \citenamefont {Koppens}}]{Chen2012}%
  \BibitemOpen
  \bibfield  {author} {\bibinfo {author} {\bibfnamefont {J.}~\bibnamefont
  {Chen}}, \bibinfo {author} {\bibfnamefont {M.}~\bibnamefont {Badioli}},
  \bibinfo {author} {\bibfnamefont {P.}~\bibnamefont {Alonso-Gonz\'{a}lez}},
  \bibinfo {author} {\bibfnamefont {S.}~\bibnamefont {Thongrattanasiri}},
  \bibinfo {author} {\bibfnamefont {F.}~\bibnamefont {Huth}}, \bibinfo {author}
  {\bibfnamefont {J.}~\bibnamefont {Osmond}}, \bibinfo {author} {\bibfnamefont
  {M.}~\bibnamefont {Spasenovi\'{c}}}, \bibinfo {author} {\bibfnamefont
  {A.}~\bibnamefont {Centeno}}, \bibinfo {author} {\bibfnamefont
  {A.}~\bibnamefont {Pesquera}}, \bibinfo {author} {\bibfnamefont
  {P.}~\bibnamefont {Godignon}}, \bibinfo {author} {\bibfnamefont
  {A.}~\bibnamefont {{Zurutuza Elorza}}}, \bibinfo {author} {\bibfnamefont
  {N.}~\bibnamefont {Camara}}, \bibinfo {author} {\bibfnamefont {F.~J.}\
  \bibnamefont {{Garc\'{\i}a de Abajo}}}, \bibinfo {author} {\bibfnamefont
  {R.}~\bibnamefont {Hillenbrand}}, \ and\ \bibinfo {author} {\bibfnamefont
  {F.~H.~L.}\ \bibnamefont {Koppens}},\ }\href {\doibase 10.1038/nature11254}
  {\bibfield  {journal} {\bibinfo  {journal} {Nature}\ }\textbf {\bibinfo
  {volume} {487}},\ \bibinfo {pages} {77} (\bibinfo {year} {2012})}\BibitemShut
  {NoStop}%
\bibitem [{\citenamefont {Alonso-Gonz\'{a}lez}\ \emph
  {et~al.}(2014)\citenamefont {Alonso-Gonz\'{a}lez}, \citenamefont {Nikitin},
  \citenamefont {Golmar}, \citenamefont {Centeno}, \citenamefont {Pesquera},
  \citenamefont {Velez}, \citenamefont {Chen}, \citenamefont {Navickaite},
  \citenamefont {Koppens}, \citenamefont {Zurutuza}, \citenamefont {Casanova},
  \citenamefont {Hueso},\ and\ \citenamefont
  {Hillenbrand}}]{Alonso-Gonzalez2014}%
  \BibitemOpen
  \bibfield  {author} {\bibinfo {author} {\bibfnamefont {P.}~\bibnamefont
  {Alonso-Gonz\'{a}lez}}, \bibinfo {author} {\bibfnamefont {A.~Y.}\
  \bibnamefont {Nikitin}}, \bibinfo {author} {\bibfnamefont {F.}~\bibnamefont
  {Golmar}}, \bibinfo {author} {\bibfnamefont {A.}~\bibnamefont {Centeno}},
  \bibinfo {author} {\bibfnamefont {A.}~\bibnamefont {Pesquera}}, \bibinfo
  {author} {\bibfnamefont {S.}~\bibnamefont {Velez}}, \bibinfo {author}
  {\bibfnamefont {J.}~\bibnamefont {Chen}}, \bibinfo {author} {\bibfnamefont
  {G.}~\bibnamefont {Navickaite}}, \bibinfo {author} {\bibfnamefont {F.~H.}\
  \bibnamefont {Koppens}}, \bibinfo {author} {\bibfnamefont {A.}~\bibnamefont
  {Zurutuza}}, \bibinfo {author} {\bibfnamefont {F.}~\bibnamefont {Casanova}},
  \bibinfo {author} {\bibfnamefont {L.~E.}\ \bibnamefont {Hueso}}, \ and\
  \bibinfo {author} {\bibfnamefont {R.}~\bibnamefont {Hillenbrand}},\ }\href
  {\doibase 10.1126/science.1253202} {\bibfield  {journal} {\bibinfo  {journal}
  {Science}\ }\textbf {\bibinfo {volume} {344}},\ \bibinfo {pages} {1369}
  (\bibinfo {year} {2014})}\BibitemShut {NoStop}%
\bibitem [{\citenamefont {Principi}\ \emph
  {et~al.}(2013{\natexlab{a}})\citenamefont {Principi}, \citenamefont
  {Vignale}, \citenamefont {Carrega},\ and\ \citenamefont
  {Polini}}]{Principi2013c}%
  \BibitemOpen
  \bibfield  {author} {\bibinfo {author} {\bibfnamefont {A.}~\bibnamefont
  {Principi}}, \bibinfo {author} {\bibfnamefont {G.}~\bibnamefont {Vignale}},
  \bibinfo {author} {\bibfnamefont {M.}~\bibnamefont {Carrega}}, \ and\
  \bibinfo {author} {\bibfnamefont {M.}~\bibnamefont {Polini}},\ }\href
  {\doibase 10.1103/PhysRevB.88.121405} {\bibfield  {journal} {\bibinfo
  {journal} {Phys. Rev. B}\ }\textbf {\bibinfo {volume} {88}},\ \bibinfo
  {pages} {121405(R)} (\bibinfo {year} {2013}{\natexlab{a}})}\BibitemShut
  {NoStop}%
\bibitem [{\citenamefont {Dean}\ \emph {et~al.}(2010)\citenamefont {Dean},
  \citenamefont {Young}, \citenamefont {Meric}, \citenamefont {Lee},
  \citenamefont {Wang}, \citenamefont {Sorgenfrei}, \citenamefont {Watanabe},
  \citenamefont {Taniguchi}, \citenamefont {Kim}, \citenamefont {Shepard},\
  and\ \citenamefont {Hone}}]{Dean2010}%
  \BibitemOpen
  \bibfield  {author} {\bibinfo {author} {\bibfnamefont {C.~R.}\ \bibnamefont
  {Dean}}, \bibinfo {author} {\bibfnamefont {A.~F.}\ \bibnamefont {Young}},
  \bibinfo {author} {\bibfnamefont {I.}~\bibnamefont {Meric}}, \bibinfo
  {author} {\bibfnamefont {C.}~\bibnamefont {Lee}}, \bibinfo {author}
  {\bibfnamefont {L.}~\bibnamefont {Wang}}, \bibinfo {author} {\bibfnamefont
  {S.}~\bibnamefont {Sorgenfrei}}, \bibinfo {author} {\bibfnamefont
  {K.}~\bibnamefont {Watanabe}}, \bibinfo {author} {\bibfnamefont
  {T.}~\bibnamefont {Taniguchi}}, \bibinfo {author} {\bibfnamefont
  {P.}~\bibnamefont {Kim}}, \bibinfo {author} {\bibfnamefont {K.~L.}\
  \bibnamefont {Shepard}}, \ and\ \bibinfo {author} {\bibfnamefont
  {J.}~\bibnamefont {Hone}},\ }\href {\doibase 10.1038/nnano.2010.172}
  {\bibfield  {journal} {\bibinfo  {journal} {Nat. Nanotechnol.}\ }\textbf
  {\bibinfo {volume} {5}},\ \bibinfo {pages} {722} (\bibinfo {year}
  {2010})}\BibitemShut {NoStop}%
\bibitem [{\citenamefont {Geim}\ and\ \citenamefont
  {Grigorieva}(2013)}]{Geim2013}%
  \BibitemOpen
  \bibfield  {author} {\bibinfo {author} {\bibfnamefont {A.~K.}\ \bibnamefont
  {Geim}}\ and\ \bibinfo {author} {\bibfnamefont {I.~V.}\ \bibnamefont
  {Grigorieva}},\ }\href {\doibase 10.1038/nature12385} {\bibfield  {journal}
  {\bibinfo  {journal} {Nature}\ }\textbf {\bibinfo {volume} {499}},\ \bibinfo
  {pages} {419} (\bibinfo {year} {2013})}\BibitemShut {NoStop}%
\bibitem [{\citenamefont {Wang}\ \emph {et~al.}(2013)\citenamefont {Wang},
  \citenamefont {Meric}, \citenamefont {Huang}, \citenamefont {Gao},
  \citenamefont {Gao}, \citenamefont {Tran}, \citenamefont {Taniguchi},
  \citenamefont {Watanabe}, \citenamefont {Campos}, \citenamefont {Muller},
  \citenamefont {Guo}, \citenamefont {Kim}, \citenamefont {Hone}, \citenamefont
  {Shepard},\ and\ \citenamefont {Dean}}]{Wang2013}%
  \BibitemOpen
  \bibfield  {author} {\bibinfo {author} {\bibfnamefont {L.}~\bibnamefont
  {Wang}}, \bibinfo {author} {\bibfnamefont {I.}~\bibnamefont {Meric}},
  \bibinfo {author} {\bibfnamefont {P.~Y.}\ \bibnamefont {Huang}}, \bibinfo
  {author} {\bibfnamefont {Q.}~\bibnamefont {Gao}}, \bibinfo {author}
  {\bibfnamefont {Y.}~\bibnamefont {Gao}}, \bibinfo {author} {\bibfnamefont
  {H.}~\bibnamefont {Tran}}, \bibinfo {author} {\bibfnamefont {T.}~\bibnamefont
  {Taniguchi}}, \bibinfo {author} {\bibfnamefont {K.}~\bibnamefont {Watanabe}},
  \bibinfo {author} {\bibfnamefont {L.~M.}\ \bibnamefont {Campos}}, \bibinfo
  {author} {\bibfnamefont {D.~A.}\ \bibnamefont {Muller}}, \bibinfo {author}
  {\bibfnamefont {J.}~\bibnamefont {Guo}}, \bibinfo {author} {\bibfnamefont
  {P.}~\bibnamefont {Kim}}, \bibinfo {author} {\bibfnamefont {J.}~\bibnamefont
  {Hone}}, \bibinfo {author} {\bibfnamefont {K.~L.}\ \bibnamefont {Shepard}}, \
  and\ \bibinfo {author} {\bibfnamefont {C.~R.}\ \bibnamefont {Dean}},\ }\href
  {\doibase 10.1126/science.1244358} {\bibfield  {journal} {\bibinfo  {journal}
  {Science}\ }\textbf {\bibinfo {volume} {342}},\ \bibinfo {pages} {614}
  (\bibinfo {year} {2013})}\BibitemShut {NoStop}%
\bibitem [{\citenamefont {Yankowitz}\ \emph {et~al.}(2012)\citenamefont
  {Yankowitz}, \citenamefont {Xue}, \citenamefont {Cormode}, \citenamefont
  {Sanchez-Yamagishi}, \citenamefont {Watanabe}, \citenamefont {Taniguchi},
  \citenamefont {Jarillo-Herrero}, \citenamefont {Jacquod},\ and\ \citenamefont
  {LeRoy}}]{Yankowitz2012}%
  \BibitemOpen
  \bibfield  {author} {\bibinfo {author} {\bibfnamefont {M.}~\bibnamefont
  {Yankowitz}}, \bibinfo {author} {\bibfnamefont {J.}~\bibnamefont {Xue}},
  \bibinfo {author} {\bibfnamefont {D.}~\bibnamefont {Cormode}}, \bibinfo
  {author} {\bibfnamefont {J.~D.}\ \bibnamefont {Sanchez-Yamagishi}}, \bibinfo
  {author} {\bibfnamefont {K.}~\bibnamefont {Watanabe}}, \bibinfo {author}
  {\bibfnamefont {T.}~\bibnamefont {Taniguchi}}, \bibinfo {author}
  {\bibfnamefont {P.}~\bibnamefont {Jarillo-Herrero}}, \bibinfo {author}
  {\bibfnamefont {P.}~\bibnamefont {Jacquod}}, \ and\ \bibinfo {author}
  {\bibfnamefont {B.~J.}\ \bibnamefont {LeRoy}},\ }\href {\doibase
  10.1038/nphys2272} {\bibfield  {journal} {\bibinfo  {journal} {Nature Phys.}\
  }\textbf {\bibinfo {volume} {8}},\ \bibinfo {pages} {382} (\bibinfo {year}
  {2012})}\BibitemShut {NoStop}%
\bibitem [{\citenamefont {Tomadin}\ \emph {et~al.}(2014)\citenamefont
  {Tomadin}, \citenamefont {Guinea},\ and\ \citenamefont
  {Polini}}]{Tomadin2014}%
  \BibitemOpen
  \bibfield  {author} {\bibinfo {author} {\bibfnamefont {A.}~\bibnamefont
  {Tomadin}}, \bibinfo {author} {\bibfnamefont {F.}~\bibnamefont {Guinea}}, \
  and\ \bibinfo {author} {\bibfnamefont {M.}~\bibnamefont {Polini}},\ }\href
  {http://arxiv.org/abs/1407.2810} {\bibfield  {journal} {\bibinfo  {journal}
  {arXiv}\ ,\ \bibinfo {pages} {1407.2810}} (\bibinfo {year} {2014})},\ \Eprint
  {http://arxiv.org/abs/1407.2810} {arXiv:1407.2810} \BibitemShut {NoStop}%
\bibitem [{\citenamefont {Dai}\ \emph {et~al.}(2014)\citenamefont {Dai},
  \citenamefont {Fei}, \citenamefont {Ma}, \citenamefont {Rodin}, \citenamefont
  {Wagner}, \citenamefont {McLeod}, \citenamefont {Liu}, \citenamefont
  {Gannett}, \citenamefont {Regan}, \citenamefont {Watanabe}, \citenamefont
  {Taniguchi}, \citenamefont {Thiemens}, \citenamefont {Dominguez},
  \citenamefont {{Castro Neto}}, \citenamefont {Zettl}, \citenamefont
  {Keilmann}, \citenamefont {Jarillo-Herrero}, \citenamefont {Fogler},\ and\
  \citenamefont {Basov}}]{Dai2014b}%
  \BibitemOpen
  \bibfield  {author} {\bibinfo {author} {\bibfnamefont {S.}~\bibnamefont
  {Dai}}, \bibinfo {author} {\bibfnamefont {Z.}~\bibnamefont {Fei}}, \bibinfo
  {author} {\bibfnamefont {Q.}~\bibnamefont {Ma}}, \bibinfo {author}
  {\bibfnamefont {A.~S.}\ \bibnamefont {Rodin}}, \bibinfo {author}
  {\bibfnamefont {M.}~\bibnamefont {Wagner}}, \bibinfo {author} {\bibfnamefont
  {A.~S.}\ \bibnamefont {McLeod}}, \bibinfo {author} {\bibfnamefont {M.~K.}\
  \bibnamefont {Liu}}, \bibinfo {author} {\bibfnamefont {W.}~\bibnamefont
  {Gannett}}, \bibinfo {author} {\bibfnamefont {W.}~\bibnamefont {Regan}},
  \bibinfo {author} {\bibfnamefont {K.}~\bibnamefont {Watanabe}}, \bibinfo
  {author} {\bibfnamefont {T.}~\bibnamefont {Taniguchi}}, \bibinfo {author}
  {\bibfnamefont {M.}~\bibnamefont {Thiemens}}, \bibinfo {author}
  {\bibfnamefont {G.}~\bibnamefont {Dominguez}}, \bibinfo {author}
  {\bibfnamefont {A.~H.}\ \bibnamefont {{Castro Neto}}}, \bibinfo {author}
  {\bibfnamefont {A.}~\bibnamefont {Zettl}}, \bibinfo {author} {\bibfnamefont
  {F.}~\bibnamefont {Keilmann}}, \bibinfo {author} {\bibfnamefont
  {P.}~\bibnamefont {Jarillo-Herrero}}, \bibinfo {author} {\bibfnamefont
  {M.~M.}\ \bibnamefont {Fogler}}, \ and\ \bibinfo {author} {\bibfnamefont
  {D.~N.}\ \bibnamefont {Basov}},\ }\href {\doibase 10.1126/science.1246833}
  {\bibfield  {journal} {\bibinfo  {journal} {Science}\ }\textbf {\bibinfo
  {volume} {343}},\ \bibinfo {pages} {1125} (\bibinfo {year}
  {2014})}\BibitemShut {NoStop}%
\bibitem [{\citenamefont {Caldwell}\ \emph {et~al.}(2014)\citenamefont
  {Caldwell}, \citenamefont {Kretinin}, \citenamefont {Chen}, \citenamefont
  {Giannini}, \citenamefont {Fogler}, \citenamefont {Francescato},
  \citenamefont {Ellis}, \citenamefont {Tischler}, \citenamefont {Woods},
  \citenamefont {Giles}, \citenamefont {Hong}, \citenamefont {Watanabe},
  \citenamefont {Taniguchi}, \citenamefont {Maier},\ and\ \citenamefont
  {Novoselov}}]{Caldwell2014}%
  \BibitemOpen
  \bibfield  {author} {\bibinfo {author} {\bibfnamefont {J.~D.}\ \bibnamefont
  {Caldwell}}, \bibinfo {author} {\bibfnamefont {A.}~\bibnamefont {Kretinin}},
  \bibinfo {author} {\bibfnamefont {Y.}~\bibnamefont {Chen}}, \bibinfo {author}
  {\bibfnamefont {V.}~\bibnamefont {Giannini}}, \bibinfo {author}
  {\bibfnamefont {M.~M.}\ \bibnamefont {Fogler}}, \bibinfo {author}
  {\bibfnamefont {Y.}~\bibnamefont {Francescato}}, \bibinfo {author}
  {\bibfnamefont {C.~T.}\ \bibnamefont {Ellis}}, \bibinfo {author}
  {\bibfnamefont {J.~G.}\ \bibnamefont {Tischler}}, \bibinfo {author}
  {\bibfnamefont {C.~R.}\ \bibnamefont {Woods}}, \bibinfo {author}
  {\bibfnamefont {A.~J.}\ \bibnamefont {Giles}}, \bibinfo {author}
  {\bibfnamefont {M.}~\bibnamefont {Hong}}, \bibinfo {author} {\bibfnamefont
  {K.}~\bibnamefont {Watanabe}}, \bibinfo {author} {\bibfnamefont
  {T.}~\bibnamefont {Taniguchi}}, \bibinfo {author} {\bibfnamefont {S.~A.}\
  \bibnamefont {Maier}}, \ and\ \bibinfo {author} {\bibfnamefont {K.~S.}\
  \bibnamefont {Novoselov}},\ }\href {http://arxiv.org/abs/1404.0494}
  {\bibfield  {journal} {\bibinfo  {journal} {arXiv}\ ,\ \bibinfo {pages}
  {1404.0494}} (\bibinfo {year} {2014})},\ \Eprint
  {http://arxiv.org/abs/1404.0494} {arXiv:1404.0494} \BibitemShut {NoStop}%
\bibitem [{\citenamefont {Brar}\ \emph {et~al.}(2014)\citenamefont {Brar},
  \citenamefont {Jang}, \citenamefont {Sherrott}, \citenamefont {Kim},
  \citenamefont {Lopez}, \citenamefont {Kim}, \citenamefont {Choi},\ and\
  \citenamefont {Atwater}}]{Brar2014}%
  \BibitemOpen
  \bibfield  {author} {\bibinfo {author} {\bibfnamefont {V.~W.}\ \bibnamefont
  {Brar}}, \bibinfo {author} {\bibfnamefont {M.~S.}\ \bibnamefont {Jang}},
  \bibinfo {author} {\bibfnamefont {M.}~\bibnamefont {Sherrott}}, \bibinfo
  {author} {\bibfnamefont {S.}~\bibnamefont {Kim}}, \bibinfo {author}
  {\bibfnamefont {J.~J.}\ \bibnamefont {Lopez}}, \bibinfo {author}
  {\bibfnamefont {L.~B.}\ \bibnamefont {Kim}}, \bibinfo {author} {\bibfnamefont
  {M.}~\bibnamefont {Choi}}, \ and\ \bibinfo {author} {\bibfnamefont
  {H.}~\bibnamefont {Atwater}},\ }\href {\doibase 10.1021/nl501096s} {\bibfield
   {journal} {\bibinfo  {journal} {Nano Lett.}\ }\textbf {\bibinfo {volume}
  {14}},\ \bibinfo {pages} {3876} (\bibinfo {year} {2014})}\BibitemShut
  {NoStop}%
\bibitem [{\citenamefont {Fei}\ \emph {et~al.}(2011)\citenamefont {Fei},
  \citenamefont {Andreev}, \citenamefont {Bao}, \citenamefont {Zhang},
  \citenamefont {{S McLeod}}, \citenamefont {Wang}, \citenamefont {Stewart},
  \citenamefont {Zhao}, \citenamefont {Dominguez}, \citenamefont {Thiemens},
  \citenamefont {Fogler}, \citenamefont {Tauber}, \citenamefont {Castro-Neto},
  \citenamefont {Lau}, \citenamefont {Keilmann},\ and\ \citenamefont
  {Basov}}]{Fei2011}%
  \BibitemOpen
  \bibfield  {author} {\bibinfo {author} {\bibfnamefont {Z.}~\bibnamefont
  {Fei}}, \bibinfo {author} {\bibfnamefont {G.~O.}\ \bibnamefont {Andreev}},
  \bibinfo {author} {\bibfnamefont {W.}~\bibnamefont {Bao}}, \bibinfo {author}
  {\bibfnamefont {L.~M.}\ \bibnamefont {Zhang}}, \bibinfo {author}
  {\bibfnamefont {A.}~\bibnamefont {{S McLeod}}}, \bibinfo {author}
  {\bibfnamefont {C.}~\bibnamefont {Wang}}, \bibinfo {author} {\bibfnamefont
  {M.~K.}\ \bibnamefont {Stewart}}, \bibinfo {author} {\bibfnamefont
  {Z.}~\bibnamefont {Zhao}}, \bibinfo {author} {\bibfnamefont {G.}~\bibnamefont
  {Dominguez}}, \bibinfo {author} {\bibfnamefont {M.}~\bibnamefont {Thiemens}},
  \bibinfo {author} {\bibfnamefont {M.~M.}\ \bibnamefont {Fogler}}, \bibinfo
  {author} {\bibfnamefont {M.~J.}\ \bibnamefont {Tauber}}, \bibinfo {author}
  {\bibfnamefont {A.~H.}\ \bibnamefont {Castro-Neto}}, \bibinfo {author}
  {\bibfnamefont {C.~N.}\ \bibnamefont {Lau}}, \bibinfo {author} {\bibfnamefont
  {F.}~\bibnamefont {Keilmann}}, \ and\ \bibinfo {author} {\bibfnamefont
  {D.~N.}\ \bibnamefont {Basov}},\ }\href {\doibase 10.1021/nl202362d}
  {\bibfield  {journal} {\bibinfo  {journal} {Nano Lett.}\ }\textbf {\bibinfo
  {volume} {11}},\ \bibinfo {pages} {4701} (\bibinfo {year}
  {2011})}\BibitemShut {NoStop}%
\bibitem [{\citenamefont {Xue}\ \emph {et~al.}(2011)\citenamefont {Xue},
  \citenamefont {Sanchez-Yamagishi}, \citenamefont {Bulmash}, \citenamefont
  {Jacquod}, \citenamefont {Deshpande}, \citenamefont {Watanabe}, \citenamefont
  {Taniguchi}, \citenamefont {Jarillo-Herrero},\ and\ \citenamefont
  {LeRoy}}]{Xue2011}%
  \BibitemOpen
  \bibfield  {author} {\bibinfo {author} {\bibfnamefont {J.}~\bibnamefont
  {Xue}}, \bibinfo {author} {\bibfnamefont {J.}~\bibnamefont
  {Sanchez-Yamagishi}}, \bibinfo {author} {\bibfnamefont {D.}~\bibnamefont
  {Bulmash}}, \bibinfo {author} {\bibfnamefont {P.}~\bibnamefont {Jacquod}},
  \bibinfo {author} {\bibfnamefont {A.}~\bibnamefont {Deshpande}}, \bibinfo
  {author} {\bibfnamefont {K.}~\bibnamefont {Watanabe}}, \bibinfo {author}
  {\bibfnamefont {T.}~\bibnamefont {Taniguchi}}, \bibinfo {author}
  {\bibfnamefont {P.}~\bibnamefont {Jarillo-Herrero}}, \ and\ \bibinfo {author}
  {\bibfnamefont {B.~J.}\ \bibnamefont {LeRoy}},\ }\href {\doibase
  10.1038/nmat2968} {\bibfield  {journal} {\bibinfo  {journal} {Nat. Mater.}\
  }\textbf {\bibinfo {volume} {10}},\ \bibinfo {pages} {282} (\bibinfo {year}
  {2011})}\BibitemShut {NoStop}%
\bibitem [{\citenamefont {Jablan}\ \emph {et~al.}(2009)\citenamefont {Jablan},
  \citenamefont {Buljan},\ and\ \citenamefont
  {Solja\v{c}i\'{c}}}]{Jablan_PRB_2009}%
  \BibitemOpen
  \bibfield  {author} {\bibinfo {author} {\bibfnamefont {M.}~\bibnamefont
  {Jablan}}, \bibinfo {author} {\bibfnamefont {H.}~\bibnamefont {Buljan}}, \
  and\ \bibinfo {author} {\bibfnamefont {M.}~\bibnamefont {Solja\v{c}i\'{c}}},\
  }\href {\doibase 10.1103/PhysRevB.80.245435} {\bibfield  {journal} {\bibinfo
  {journal} {Phys. Rev. B}\ }\textbf {\bibinfo {volume} {80}},\ \bibinfo
  {pages} {245435} (\bibinfo {year} {2009})}\BibitemShut {NoStop}%
\bibitem [{\citenamefont {Zhang}\ \emph {et~al.}(2014)\citenamefont {Zhang},
  \citenamefont {Fu},\ and\ \citenamefont {Yang}}]{Zhang2014a}%
  \BibitemOpen
  \bibfield  {author} {\bibinfo {author} {\bibfnamefont {L.}~\bibnamefont
  {Zhang}}, \bibinfo {author} {\bibfnamefont {X.}~\bibnamefont {Fu}}, \ and\
  \bibinfo {author} {\bibfnamefont {J.}~\bibnamefont {Yang}},\ }\href
  {http://ctp.itp.ac.cn/EN/article/downloadArticleFile.do?attachType=PDF\&id=16316}
  {\bibfield  {journal} {\bibinfo  {journal} {Commun. Theor. Phys.}\ }\textbf
  {\bibinfo {volume} {61}},\ \bibinfo {pages} {751} (\bibinfo {year}
  {2014})}\BibitemShut {NoStop}%
\bibitem [{\citenamefont {Johnson}\ and\ \citenamefont
  {Christy}(1972)}]{Johnson1972}%
  \BibitemOpen
  \bibfield  {author} {\bibinfo {author} {\bibfnamefont {P.}~\bibnamefont
  {Johnson}}\ and\ \bibinfo {author} {\bibfnamefont {R.}~\bibnamefont
  {Christy}},\ }\href {\doibase 10.1103/PhysRevB.6.4370} {\bibfield  {journal}
  {\bibinfo  {journal} {Phys. Rev. B}\ }\textbf {\bibinfo {volume} {6}},\
  \bibinfo {pages} {4370} (\bibinfo {year} {1972})}\BibitemShut {NoStop}%
\bibitem [{\citenamefont {Principi}\ \emph {et~al.}(2014)\citenamefont
  {Principi}, \citenamefont {Carrega}, \citenamefont {Lundeberg}, \citenamefont
  {Woessner}, \citenamefont {Koppens}, \citenamefont {Vignale},\ and\
  \citenamefont {Polini}}]{Principi}%
  \BibitemOpen
  \bibfield  {author} {\bibinfo {author} {\bibfnamefont {A.}~\bibnamefont
  {Principi}}, \bibinfo {author} {\bibfnamefont {M.}~\bibnamefont {Carrega}},
  \bibinfo {author} {\bibfnamefont {M.}~\bibnamefont {Lundeberg}}, \bibinfo
  {author} {\bibfnamefont {A.}~\bibnamefont {Woessner}}, \bibinfo {author}
  {\bibfnamefont {F.~H.~L.}\ \bibnamefont {Koppens}}, \bibinfo {author}
  {\bibfnamefont {G.}~\bibnamefont {Vignale}}, \ and\ \bibinfo {author}
  {\bibfnamefont {M.}~\bibnamefont {Polini}},\ }\href
  {http://arxiv.org/abs/1408.1653} {\bibfield  {journal} {\bibinfo  {journal}
  {arXiv}\ ,\ \bibinfo {pages} {1408.1653}} (\bibinfo {year} {2014})},\ \Eprint
  {http://arxiv.org/abs/1408.1653} {arXiv:1408.1653} \BibitemShut {NoStop}%
\bibitem [{\citenamefont {Yan}\ \emph {et~al.}(2013)\citenamefont {Yan},
  \citenamefont {Low}, \citenamefont {Zhu}, \citenamefont {Wu}, \citenamefont
  {Freitag}, \citenamefont {Li}, \citenamefont {Guinea}, \citenamefont
  {Avouris},\ and\ \citenamefont {Xia}}]{Yan2013c}%
  \BibitemOpen
  \bibfield  {author} {\bibinfo {author} {\bibfnamefont {H.}~\bibnamefont
  {Yan}}, \bibinfo {author} {\bibfnamefont {T.}~\bibnamefont {Low}}, \bibinfo
  {author} {\bibfnamefont {W.}~\bibnamefont {Zhu}}, \bibinfo {author}
  {\bibfnamefont {Y.}~\bibnamefont {Wu}}, \bibinfo {author} {\bibfnamefont
  {M.}~\bibnamefont {Freitag}}, \bibinfo {author} {\bibfnamefont
  {X.}~\bibnamefont {Li}}, \bibinfo {author} {\bibfnamefont {F.}~\bibnamefont
  {Guinea}}, \bibinfo {author} {\bibfnamefont {P.}~\bibnamefont {Avouris}}, \
  and\ \bibinfo {author} {\bibfnamefont {F.}~\bibnamefont {Xia}},\ }\href
  {\doibase 10.1038/nphoton.2013.57} {\bibfield  {journal} {\bibinfo  {journal}
  {Nature Photon.}\ }\textbf {\bibinfo {volume} {7}},\ \bibinfo {pages} {394}
  (\bibinfo {year} {2013})}\BibitemShut {NoStop}%
\bibitem [{\citenamefont {Principi}\ \emph
  {et~al.}(2013{\natexlab{b}})\citenamefont {Principi}, \citenamefont
  {Vignale}, \citenamefont {Carrega},\ and\ \citenamefont
  {Polini}}]{Principi2013b}%
  \BibitemOpen
  \bibfield  {author} {\bibinfo {author} {\bibfnamefont {A.}~\bibnamefont
  {Principi}}, \bibinfo {author} {\bibfnamefont {G.}~\bibnamefont {Vignale}},
  \bibinfo {author} {\bibfnamefont {M.}~\bibnamefont {Carrega}}, \ and\
  \bibinfo {author} {\bibfnamefont {M.}~\bibnamefont {Polini}},\ }\href
  {\doibase 10.1103/PhysRevB.88.195405} {\bibfield  {journal} {\bibinfo
  {journal} {Phys. Rev. B}\ }\textbf {\bibinfo {volume} {88}},\ \bibinfo
  {pages} {195405} (\bibinfo {year} {2013}{\natexlab{b}})}\BibitemShut
  {NoStop}%
\bibitem [{\citenamefont {Li}\ \emph {et~al.}(2008)\citenamefont {Li},
  \citenamefont {Henriksen}, \citenamefont {Jiang}, \citenamefont {Hao},
  \citenamefont {Martin}, \citenamefont {Kim}, \citenamefont {Stormer},\ and\
  \citenamefont {Basov}}]{Li2008f}%
  \BibitemOpen
  \bibfield  {author} {\bibinfo {author} {\bibfnamefont {Z.~Q.}\ \bibnamefont
  {Li}}, \bibinfo {author} {\bibfnamefont {E.~A.}\ \bibnamefont {Henriksen}},
  \bibinfo {author} {\bibfnamefont {Z.}~\bibnamefont {Jiang}}, \bibinfo
  {author} {\bibfnamefont {Z.}~\bibnamefont {Hao}}, \bibinfo {author}
  {\bibfnamefont {M.~C.}\ \bibnamefont {Martin}}, \bibinfo {author}
  {\bibfnamefont {P.}~\bibnamefont {Kim}}, \bibinfo {author} {\bibfnamefont
  {H.~L.}\ \bibnamefont {Stormer}}, \ and\ \bibinfo {author} {\bibfnamefont
  {D.~N.}\ \bibnamefont {Basov}},\ }\href {\doibase 10.1038/nphys989}
  {\bibfield  {journal} {\bibinfo  {journal} {Nature Phys.}\ }\textbf {\bibinfo
  {volume} {4}},\ \bibinfo {pages} {532} (\bibinfo {year} {2008})}\BibitemShut
  {NoStop}%
\bibitem [{\citenamefont {Mak}\ \emph {et~al.}(2008)\citenamefont {Mak},
  \citenamefont {Sfeir}, \citenamefont {Wu}, \citenamefont {Lui}, \citenamefont
  {Misewich},\ and\ \citenamefont {Heinz}}]{Mak2008b}%
  \BibitemOpen
  \bibfield  {author} {\bibinfo {author} {\bibfnamefont {K.~F.}\ \bibnamefont
  {Mak}}, \bibinfo {author} {\bibfnamefont {M.~Y.}\ \bibnamefont {Sfeir}},
  \bibinfo {author} {\bibfnamefont {Y.}~\bibnamefont {Wu}}, \bibinfo {author}
  {\bibfnamefont {C.~H.}\ \bibnamefont {Lui}}, \bibinfo {author} {\bibfnamefont
  {J.~A.}\ \bibnamefont {Misewich}}, \ and\ \bibinfo {author} {\bibfnamefont
  {T.~F.}\ \bibnamefont {Heinz}},\ }\href {\doibase
  10.1103/PhysRevLett.101.196405} {\bibfield  {journal} {\bibinfo  {journal}
  {Phys. Rev. Lett.}\ }\textbf {\bibinfo {volume} {101}},\ \bibinfo {pages}
  {196405} (\bibinfo {year} {2008})}\BibitemShut {NoStop}%
\bibitem [{\citenamefont {Christensen}\ \emph {et~al.}(2012)\citenamefont
  {Christensen}, \citenamefont {Manjavacas}, \citenamefont {Thongrattanasiri},
  \citenamefont {Koppens},\ and\ \citenamefont {{Garc\'{\i}a de
  Abajo}}}]{Christensen2012}%
  \BibitemOpen
  \bibfield  {author} {\bibinfo {author} {\bibfnamefont {J.}~\bibnamefont
  {Christensen}}, \bibinfo {author} {\bibfnamefont {A.}~\bibnamefont
  {Manjavacas}}, \bibinfo {author} {\bibfnamefont {S.}~\bibnamefont
  {Thongrattanasiri}}, \bibinfo {author} {\bibfnamefont {F.~H.~L.}\
  \bibnamefont {Koppens}}, \ and\ \bibinfo {author} {\bibfnamefont {F.~J.}\
  \bibnamefont {{Garc\'{\i}a de Abajo}}},\ }\href {\doibase 10.1021/nn2037626}
  {\bibfield  {journal} {\bibinfo  {journal} {ACS Nano}\ }\textbf {\bibinfo
  {volume} {6}},\ \bibinfo {pages} {431} (\bibinfo {year} {2012})}\BibitemShut
  {NoStop}%
\bibitem [{\citenamefont {Vakil}\ and\ \citenamefont
  {Engheta}(2011)}]{Vakil2011}%
  \BibitemOpen
  \bibfield  {author} {\bibinfo {author} {\bibfnamefont {A.}~\bibnamefont
  {Vakil}}\ and\ \bibinfo {author} {\bibfnamefont {N.}~\bibnamefont
  {Engheta}},\ }\href {\doibase 10.1126/science.1202691} {\bibfield  {journal}
  {\bibinfo  {journal} {Science}\ }\textbf {\bibinfo {volume} {332}},\ \bibinfo
  {pages} {1291} (\bibinfo {year} {2011})}\BibitemShut {NoStop}%
\bibitem [{\citenamefont {Wunsch}\ \emph {et~al.}(2006)\citenamefont {Wunsch},
  \citenamefont {Stauber}, \citenamefont {Sols},\ and\ \citenamefont
  {Guinea}}]{Wunsch2006}%
  \BibitemOpen
  \bibfield  {author} {\bibinfo {author} {\bibfnamefont {B.}~\bibnamefont
  {Wunsch}}, \bibinfo {author} {\bibfnamefont {T.}~\bibnamefont {Stauber}},
  \bibinfo {author} {\bibfnamefont {F.}~\bibnamefont {Sols}}, \ and\ \bibinfo
  {author} {\bibfnamefont {F.}~\bibnamefont {Guinea}},\ }\href {\doibase
  10.1088/1367-2630/8/12/318} {\bibfield  {journal} {\bibinfo  {journal} {New
  J. Phys.}\ }\textbf {\bibinfo {volume} {8}},\ \bibinfo {pages} {318}
  (\bibinfo {year} {2006})}\BibitemShut {NoStop}%
\bibitem [{\citenamefont {Hwang}\ and\ \citenamefont {{Das
  Sarma}}(2007)}]{Hwang2007}%
  \BibitemOpen
  \bibfield  {author} {\bibinfo {author} {\bibfnamefont {E.~H.}\ \bibnamefont
  {Hwang}}\ and\ \bibinfo {author} {\bibfnamefont {S.}~\bibnamefont {{Das
  Sarma}}},\ }\href {\doibase 10.1103/PhysRevB.75.205418} {\bibfield  {journal}
  {\bibinfo  {journal} {Phys. Rev. B}\ }\textbf {\bibinfo {volume} {75}},\
  \bibinfo {pages} {205418} (\bibinfo {year} {2007})}\BibitemShut {NoStop}%
\bibitem [{\citenamefont {Principi}\ \emph {et~al.}(2009)\citenamefont
  {Principi}, \citenamefont {Polini},\ and\ \citenamefont
  {Vignale}}]{Principi2009a}%
  \BibitemOpen
  \bibfield  {author} {\bibinfo {author} {\bibfnamefont {A.}~\bibnamefont
  {Principi}}, \bibinfo {author} {\bibfnamefont {M.}~\bibnamefont {Polini}}, \
  and\ \bibinfo {author} {\bibfnamefont {G.}~\bibnamefont {Vignale}},\ }\href
  {\doibase 10.1103/PhysRevB.80.075418} {\bibfield  {journal} {\bibinfo
  {journal} {Phys. Rev. B}\ }\textbf {\bibinfo {volume} {80}},\ \bibinfo
  {pages} {075418} (\bibinfo {year} {2009})}\BibitemShut {NoStop}%
\bibitem [{\citenamefont {Cai}\ \emph {et~al.}(2007)\citenamefont {Cai},
  \citenamefont {Zhang}, \citenamefont {Zeng}, \citenamefont {Cheng},\ and\
  \citenamefont {Xu}}]{Cai2007}%
  \BibitemOpen
  \bibfield  {author} {\bibinfo {author} {\bibfnamefont {Y.}~\bibnamefont
  {Cai}}, \bibinfo {author} {\bibfnamefont {L.}~\bibnamefont {Zhang}}, \bibinfo
  {author} {\bibfnamefont {Q.}~\bibnamefont {Zeng}}, \bibinfo {author}
  {\bibfnamefont {L.}~\bibnamefont {Cheng}}, \ and\ \bibinfo {author}
  {\bibfnamefont {Y.}~\bibnamefont {Xu}},\ }\href {\doibase
  10.1016/j.ssc.2006.10.040} {\bibfield  {journal} {\bibinfo  {journal} {Solid
  State Commun.}\ }\textbf {\bibinfo {volume} {141}},\ \bibinfo {pages} {262}
  (\bibinfo {year} {2007})}\BibitemShut {NoStop}%
\end{thebibliography}

\begin{thebibliography}{16}%
\makeatletter
\providecommand \@ifxundefined [1]{%
 \@ifx{#1\undefined}
}%
\providecommand \@ifnum [1]{%
 \ifnum #1\expandafter \@firstoftwo
 \else \expandafter \@secondoftwo
 \fi
}%
\providecommand \@ifx [1]{%
 \ifx #1\expandafter \@firstoftwo
 \else \expandafter \@secondoftwo
 \fi
}%
\providecommand \natexlab [1]{#1}%
\providecommand \enquote  [1]{``#1''}%
\providecommand \bibnamefont  [1]{#1}%
\providecommand \bibfnamefont [1]{#1}%
\providecommand \citenamefont [1]{#1}%
\providecommand \href@noop [0]{\@secondoftwo}%
\providecommand \href [0]{\begingroup \@sanitize@url \@href}%
\providecommand \@href[1]{\@@startlink{#1}\@@href}%
\providecommand \@@href[1]{\endgroup#1\@@endlink}%
\providecommand \@sanitize@url [0]{\catcode `\\12\catcode `\$12\catcode
  `\&12\catcode `\#12\catcode `\^12\catcode `\_12\catcode `\%12\relax}%
\providecommand \@@startlink[1]{}%
\providecommand \@@endlink[0]{}%
\providecommand \url  [0]{\begingroup\@sanitize@url \@url }%
\providecommand \@url [1]{\endgroup\@href {#1}{\urlprefix }}%
\providecommand \urlprefix  [0]{URL }%
\providecommand \Eprint [0]{\href }%
\providecommand \doibase [0]{http://dx.doi.org/}%
\providecommand \selectlanguage [0]{\@gobble}%
\providecommand \bibinfo  [0]{\@secondoftwo}%
\providecommand \bibfield  [0]{\@secondoftwo}%
\providecommand \translation [1]{[#1]}%
\providecommand \BibitemOpen [0]{}%
\providecommand \bibitemStop [0]{}%
\providecommand \bibitemNoStop [0]{.\EOS\space}%
\providecommand \EOS [0]{\spacefactor3000\relax}%
\providecommand \BibitemShut  [1]{\csname bibitem#1\endcsname}%
\let\auto@bib@innerbib\@empty
\bibitem [{\citenamefont {Fei}\ \emph {et~al.}(2011)\citenamefont {Fei},
  \citenamefont {Andreev}, \citenamefont {Bao}, \citenamefont {Zhang},
  \citenamefont {{S McLeod}}, \citenamefont {Wang}, \citenamefont {Stewart},
  \citenamefont {Zhao}, \citenamefont {Dominguez}, \citenamefont {Thiemens},
  \citenamefont {Fogler}, \citenamefont {Tauber}, \citenamefont {Castro-Neto},
  \citenamefont {Lau}, \citenamefont {Keilmann},\ and\ \citenamefont
  {Basov}}]{SFei2011}%
  \BibitemOpen
  \bibfield  {author} {\bibinfo {author} {\bibfnamefont {Z.}~\bibnamefont
  {Fei}}, \bibinfo {author} {\bibfnamefont {G.~O.}\ \bibnamefont {Andreev}},
  \bibinfo {author} {\bibfnamefont {W.}~\bibnamefont {Bao}}, \bibinfo {author}
  {\bibfnamefont {L.~M.}\ \bibnamefont {Zhang}}, \bibinfo {author}
  {\bibfnamefont {A.}~\bibnamefont {{S McLeod}}}, \bibinfo {author}
  {\bibfnamefont {C.}~\bibnamefont {Wang}}, \bibinfo {author} {\bibfnamefont
  {M.~K.}\ \bibnamefont {Stewart}}, \bibinfo {author} {\bibfnamefont
  {Z.}~\bibnamefont {Zhao}}, \bibinfo {author} {\bibfnamefont {G.}~\bibnamefont
  {Dominguez}}, \bibinfo {author} {\bibfnamefont {M.}~\bibnamefont {Thiemens}},
  \bibinfo {author} {\bibfnamefont {M.~M.}\ \bibnamefont {Fogler}}, \bibinfo
  {author} {\bibfnamefont {M.~J.}\ \bibnamefont {Tauber}}, \bibinfo {author}
  {\bibfnamefont {A.~H.}\ \bibnamefont {Castro-Neto}}, \bibinfo {author}
  {\bibfnamefont {C.~N.}\ \bibnamefont {Lau}}, \bibinfo {author} {\bibfnamefont
  {F.}~\bibnamefont {Keilmann}}, \ and\ \bibinfo {author} {\bibfnamefont
  {D.~N.}\ \bibnamefont {Basov}},\ }\href {\doibase 10.1021/nl202362d}
  {\bibfield  {journal} {\bibinfo  {journal} {Nano Lett.}\ }\textbf {\bibinfo
  {volume} {11}},\ \bibinfo {pages} {4701} (\bibinfo {year}
  {2011})}\BibitemShut {NoStop}%
\bibitem [{\citenamefont {Nikitin}\ \emph {et~al.}(2014)\citenamefont
  {Nikitin}, \citenamefont {Low},\ and\ \citenamefont
  {Martin-Moreno}}]{SNikitin2014b}%
  \BibitemOpen
  \bibfield  {author} {\bibinfo {author} {\bibfnamefont {A.~Y.}\ \bibnamefont
  {Nikitin}}, \bibinfo {author} {\bibfnamefont {T.}~\bibnamefont {Low}}, \ and\
  \bibinfo {author} {\bibfnamefont {L.}~\bibnamefont {Martin-Moreno}},\ }\href
  {\doibase 10.1103/PhysRevB.90.041407} {\bibfield  {journal} {\bibinfo
  {journal} {Phys. Rev. B}\ }\textbf {\bibinfo {volume} {90}},\ \bibinfo
  {pages} {041407} (\bibinfo {year} {2014})}\BibitemShut {NoStop}%
\bibitem [{\citenamefont {Ocelic}\ \emph {et~al.}(2006)\citenamefont {Ocelic},
  \citenamefont {Huber},\ and\ \citenamefont {Hillenbrand}}]{SOcelic2006}%
  \BibitemOpen
  \bibfield  {author} {\bibinfo {author} {\bibfnamefont {N.}~\bibnamefont
  {Ocelic}}, \bibinfo {author} {\bibfnamefont {A.}~\bibnamefont {Huber}}, \
  and\ \bibinfo {author} {\bibfnamefont {R.}~\bibnamefont {Hillenbrand}},\
  }\href {\doibase 10.1063/1.2348781} {\bibfield  {journal} {\bibinfo
  {journal} {Appl. Phys. Lett.}\ }\textbf {\bibinfo {volume} {89}},\ \bibinfo
  {pages} {101124} (\bibinfo {year} {2006})}\BibitemShut {NoStop}%
\bibitem [{\citenamefont {Zhang}\ \emph {et~al.}(2014)\citenamefont {Zhang},
  \citenamefont {Fu},\ and\ \citenamefont {Yang}}]{SZhang2014a}%
  \BibitemOpen
  \bibfield  {author} {\bibinfo {author} {\bibfnamefont {L.}~\bibnamefont
  {Zhang}}, \bibinfo {author} {\bibfnamefont {X.}~\bibnamefont {Fu}}, \ and\
  \bibinfo {author} {\bibfnamefont {J.}~\bibnamefont {Yang}},\ }\href
  {http://ctp.itp.ac.cn/EN/article/downloadArticleFile.do?attachType=PDF\&id=16316}
  {\bibfield  {journal} {\bibinfo  {journal} {Commun. Theor. Phys.}\ }\textbf
  {\bibinfo {volume} {61}},\ \bibinfo {pages} {751} (\bibinfo {year}
  {2014})}\BibitemShut {NoStop}%
\bibitem [{\citenamefont {Craeye}\ \emph {et~al.}(1999)\citenamefont {Craeye},
  \citenamefont {Sobieski}, \citenamefont {Bliven},\ and\ \citenamefont
  {Guissard}}]{SCraeye1999}%
  \BibitemOpen
  \bibfield  {author} {\bibinfo {author} {\bibfnamefont {C.}~\bibnamefont
  {Craeye}}, \bibinfo {author} {\bibfnamefont {P.}~\bibnamefont {Sobieski}},
  \bibinfo {author} {\bibfnamefont {L.}~\bibnamefont {Bliven}}, \ and\ \bibinfo
  {author} {\bibfnamefont {A.}~\bibnamefont {Guissard}},\ }\href {\doibase
  10.1109/48.775294} {\bibfield  {journal} {\bibinfo  {journal} {IEEE J. Ocean.
  Eng}\ }\textbf {\bibinfo {volume} {24}},\ \bibinfo {pages} {323} (\bibinfo
  {year} {1999})}\BibitemShut {NoStop}%
\bibitem [{\citenamefont {Wengrovitz}\ \emph {et~al.}(1987)\citenamefont
  {Wengrovitz}, \citenamefont {Oppenheim},\ and\ \citenamefont
  {Frisk}}]{SWengrovitz1987}%
  \BibitemOpen
  \bibfield  {author} {\bibinfo {author} {\bibfnamefont {M.~S.}\ \bibnamefont
  {Wengrovitz}}, \bibinfo {author} {\bibfnamefont {A.~V.}\ \bibnamefont
  {Oppenheim}}, \ and\ \bibinfo {author} {\bibfnamefont {G.~V.}\ \bibnamefont
  {Frisk}},\ }\href {\doibase 10.1364/JOSAA.4.000247} {\bibfield  {journal}
  {\bibinfo  {journal} {J. Opt. Soc. Am. A}\ }\textbf {\bibinfo {volume} {4}},\
  \bibinfo {pages} {247} (\bibinfo {year} {1987})}\BibitemShut {NoStop}%
\bibitem [{\citenamefont {Koppens}\ \emph {et~al.}(2011)\citenamefont
  {Koppens}, \citenamefont {Chang},\ and\ \citenamefont {{Garc\'{\i}a de
  Abajo}}}]{SKoppens_Nano_Lett_2011}%
  \BibitemOpen
  \bibfield  {author} {\bibinfo {author} {\bibfnamefont {F.~H.~L.}\
  \bibnamefont {Koppens}}, \bibinfo {author} {\bibfnamefont {D.~E.}\
  \bibnamefont {Chang}}, \ and\ \bibinfo {author} {\bibfnamefont {F.~J.}\
  \bibnamefont {{Garc\'{\i}a de Abajo}}},\ }\href {\doibase 10.1021/nl201771h}
  {\bibfield  {journal} {\bibinfo  {journal} {Nano Lett.}\ }\textbf {\bibinfo
  {volume} {11}},\ \bibinfo {pages} {3370} (\bibinfo {year}
  {2011})}\BibitemShut {NoStop}%
\bibitem [{\citenamefont {Wunsch}\ \emph {et~al.}(2006)\citenamefont {Wunsch},
  \citenamefont {Stauber}, \citenamefont {Sols},\ and\ \citenamefont
  {Guinea}}]{SWunsch2006}%
  \BibitemOpen
  \bibfield  {author} {\bibinfo {author} {\bibfnamefont {B.}~\bibnamefont
  {Wunsch}}, \bibinfo {author} {\bibfnamefont {T.}~\bibnamefont {Stauber}},
  \bibinfo {author} {\bibfnamefont {F.}~\bibnamefont {Sols}}, \ and\ \bibinfo
  {author} {\bibfnamefont {F.}~\bibnamefont {Guinea}},\ }\href {\doibase
  10.1088/1367-2630/8/12/318} {\bibfield  {journal} {\bibinfo  {journal} {New
  J. Phys.}\ }\textbf {\bibinfo {volume} {8}},\ \bibinfo {pages} {318}
  (\bibinfo {year} {2006})}\BibitemShut {NoStop}%
\bibitem [{\citenamefont {Hwang}\ and\ \citenamefont {{Das
  Sarma}}(2007)}]{SHwang2007}%
  \BibitemOpen
  \bibfield  {author} {\bibinfo {author} {\bibfnamefont {E.~H.}\ \bibnamefont
  {Hwang}}\ and\ \bibinfo {author} {\bibfnamefont {S.}~\bibnamefont {{Das
  Sarma}}},\ }\href {\doibase 10.1103/PhysRevB.75.205418} {\bibfield  {journal}
  {\bibinfo  {journal} {Phys. Rev. B}\ }\textbf {\bibinfo {volume} {75}},\
  \bibinfo {pages} {205418} (\bibinfo {year} {2007})}\BibitemShut {NoStop}%
\bibitem [{\citenamefont {Jablan}\ \emph {et~al.}(2009)\citenamefont {Jablan},
  \citenamefont {Buljan},\ and\ \citenamefont
  {Solja\v{c}i\'{c}}}]{SJablan_PRB_2009}%
  \BibitemOpen
  \bibfield  {author} {\bibinfo {author} {\bibfnamefont {M.}~\bibnamefont
  {Jablan}}, \bibinfo {author} {\bibfnamefont {H.}~\bibnamefont {Buljan}}, \
  and\ \bibinfo {author} {\bibfnamefont {M.}~\bibnamefont {Solja\v{c}i\'{c}}},\
  }\href {\doibase 10.1103/PhysRevB.80.245435} {\bibfield  {journal} {\bibinfo
  {journal} {Phys. Rev. B}\ }\textbf {\bibinfo {volume} {80}},\ \bibinfo
  {pages} {245435} (\bibinfo {year} {2009})}\BibitemShut {NoStop}%
\bibitem [{\citenamefont {SPrincipi}\ \emph {et~al.}(2014)\citenamefont
  {SPrincipi}, \citenamefont {Carrega}, \citenamefont {Lundeberg}, \citenamefont
  {Woessner}, \citenamefont {Koppens}, \citenamefont {Vignale},\ and\
  \citenamefont {Polini}}]{SPrincipi}%
  \BibitemOpen
  \bibfield  {author} {\bibinfo {author} {\bibfnamefont {A.}~\bibnamefont
  {SPrincipi}}, \bibinfo {author} {\bibfnamefont {M.}~\bibnamefont {Carrega}},
  \bibinfo {author} {\bibfnamefont {M.}~\bibnamefont {Lundeberg}}, \bibinfo
  {author} {\bibfnamefont {A.}~\bibnamefont {Woessner}}, \bibinfo {author}
  {\bibfnamefont {F.~H.~L.}\ \bibnamefont {Koppens}}, \bibinfo {author}
  {\bibfnamefont {G.}~\bibnamefont {Vignale}}, \ and\ \bibinfo {author}
  {\bibfnamefont {M.}~\bibnamefont {Polini}},\ }\href
  {http://arxiv.org/abs/1408.1653} {\bibfield  {journal} {\bibinfo  {journal}
  {arXiv}\ ,\ \bibinfo {pages} {1408.1653}} (\bibinfo {year} {2014})},\ \Eprint
  {http://arxiv.org/abs/1408.1653} {arXiv:1408.1653} \BibitemShut {NoStop}%
\bibitem [{\citenamefont {Geick}\ \emph {et~al.}(1966)\citenamefont {Geick},
  \citenamefont {Perry},\ and\ \citenamefont {Rupprecht}}]{SGeick1966a}%
  \BibitemOpen
  \bibfield  {author} {\bibinfo {author} {\bibfnamefont {R.}~\bibnamefont
  {Geick}}, \bibinfo {author} {\bibfnamefont {C.}~\bibnamefont {Perry}}, \ and\
  \bibinfo {author} {\bibfnamefont {G.}~\bibnamefont {Rupprecht}},\ }\href
  {\doibase 10.1103/PhysRev.146.543} {\bibfield  {journal} {\bibinfo  {journal}
  {Phys. Rev.}\ }\textbf {\bibinfo {volume} {146}},\ \bibinfo {pages} {543}
  (\bibinfo {year} {1966})}\BibitemShut {NoStop}%
\bibitem [{\citenamefont {Caldwell}\ \emph {et~al.}(2014)\citenamefont
  {Caldwell}, \citenamefont {Kretinin}, \citenamefont {Chen}, \citenamefont
  {Giannini}, \citenamefont {Fogler}, \citenamefont {Francescato},
  \citenamefont {Ellis}, \citenamefont {Tischler}, \citenamefont {Woods},
  \citenamefont {Giles}, \citenamefont {Hong}, \citenamefont {Watanabe},
  \citenamefont {Taniguchi}, \citenamefont {Maier},\ and\ \citenamefont
  {Novoselov}}]{SCaldwell2014}%
  \BibitemOpen
  \bibfield  {author} {\bibinfo {author} {\bibfnamefont {J.~D.}\ \bibnamefont
  {Caldwell}}, \bibinfo {author} {\bibfnamefont {A.}~\bibnamefont {Kretinin}},
  \bibinfo {author} {\bibfnamefont {Y.}~\bibnamefont {Chen}}, \bibinfo {author}
  {\bibfnamefont {V.}~\bibnamefont {Giannini}}, \bibinfo {author}
  {\bibfnamefont {M.~M.}\ \bibnamefont {Fogler}}, \bibinfo {author}
  {\bibfnamefont {Y.}~\bibnamefont {Francescato}}, \bibinfo {author}
  {\bibfnamefont {C.~T.}\ \bibnamefont {Ellis}}, \bibinfo {author}
  {\bibfnamefont {J.~G.}\ \bibnamefont {Tischler}}, \bibinfo {author}
  {\bibfnamefont {C.~R.}\ \bibnamefont {Woods}}, \bibinfo {author}
  {\bibfnamefont {A.~J.}\ \bibnamefont {Giles}}, \bibinfo {author}
  {\bibfnamefont {M.}~\bibnamefont {Hong}}, \bibinfo {author} {\bibfnamefont
  {K.}~\bibnamefont {Watanabe}}, \bibinfo {author} {\bibfnamefont
  {T.}~\bibnamefont {Taniguchi}}, \bibinfo {author} {\bibfnamefont {S.~A.}\
  \bibnamefont {Maier}}, \ and\ \bibinfo {author} {\bibfnamefont {K.~S.}\
  \bibnamefont {Novoselov}},\ }\href {http://arxiv.org/abs/1404.0494}
  {\bibfield  {journal} {\bibinfo  {journal} {arXiv}\ ,\ \bibinfo {pages}
  {1404.0494}} (\bibinfo {year} {2014})},\ \Eprint
  {http://arxiv.org/abs/1404.0494} {arXiv:1404.0494} \BibitemShut {NoStop}%
\bibitem [{\citenamefont {Dai}\ \emph {et~al.}(2014)\citenamefont {Dai},
  \citenamefont {Fei}, \citenamefont {Ma}, \citenamefont {Rodin}, \citenamefont
  {Wagner}, \citenamefont {McLeod}, \citenamefont {Liu}, \citenamefont
  {Gannett}, \citenamefont {Regan}, \citenamefont {Watanabe}, \citenamefont
  {Taniguchi}, \citenamefont {Thiemens}, \citenamefont {Dominguez},
  \citenamefont {{Castro Neto}}, \citenamefont {Zettl}, \citenamefont
  {Keilmann}, \citenamefont {Jarillo-Herrero}, \citenamefont {Fogler},\ and\
  \citenamefont {Basov}}]{SDai2014b}%
  \BibitemOpen
  \bibfield  {author} {\bibinfo {author} {\bibfnamefont {S.}~\bibnamefont
  {Dai}}, \bibinfo {author} {\bibfnamefont {Z.}~\bibnamefont {Fei}}, \bibinfo
  {author} {\bibfnamefont {Q.}~\bibnamefont {Ma}}, \bibinfo {author}
  {\bibfnamefont {A.~S.}\ \bibnamefont {Rodin}}, \bibinfo {author}
  {\bibfnamefont {M.}~\bibnamefont {Wagner}}, \bibinfo {author} {\bibfnamefont
  {A.~S.}\ \bibnamefont {McLeod}}, \bibinfo {author} {\bibfnamefont {M.~K.}\
  \bibnamefont {Liu}}, \bibinfo {author} {\bibfnamefont {W.}~\bibnamefont
  {Gannett}}, \bibinfo {author} {\bibfnamefont {W.}~\bibnamefont {Regan}},
  \bibinfo {author} {\bibfnamefont {K.}~\bibnamefont {Watanabe}}, \bibinfo
  {author} {\bibfnamefont {T.}~\bibnamefont {Taniguchi}}, \bibinfo {author}
  {\bibfnamefont {M.}~\bibnamefont {Thiemens}}, \bibinfo {author}
  {\bibfnamefont {G.}~\bibnamefont {Dominguez}}, \bibinfo {author}
  {\bibfnamefont {A.~H.}\ \bibnamefont {{Castro Neto}}}, \bibinfo {author}
  {\bibfnamefont {A.}~\bibnamefont {Zettl}}, \bibinfo {author} {\bibfnamefont
  {F.}~\bibnamefont {Keilmann}}, \bibinfo {author} {\bibfnamefont
  {P.}~\bibnamefont {Jarillo-Herrero}}, \bibinfo {author} {\bibfnamefont
  {M.~M.}\ \bibnamefont {Fogler}}, \ and\ \bibinfo {author} {\bibfnamefont
  {D.~N.}\ \bibnamefont {Basov}},\ }\href {\doibase 10.1126/science.1246833}
  {\bibfield  {journal} {\bibinfo  {journal} {Science}\ }\textbf {\bibinfo
  {volume} {343}},\ \bibinfo {pages} {1125} (\bibinfo {year}
  {2014})}\BibitemShut {NoStop}%
\bibitem [{\citenamefont {Cai}\ \emph {et~al.}(2007)\citenamefont {Cai},
  \citenamefont {Zhang}, \citenamefont {Zeng}, \citenamefont {Cheng},\ and\
  \citenamefont {Xu}}]{SCai2007}%
  \BibitemOpen
  \bibfield  {author} {\bibinfo {author} {\bibfnamefont {Y.}~\bibnamefont
  {Cai}}, \bibinfo {author} {\bibfnamefont {L.}~\bibnamefont {Zhang}}, \bibinfo
  {author} {\bibfnamefont {Q.}~\bibnamefont {Zeng}}, \bibinfo {author}
  {\bibfnamefont {L.}~\bibnamefont {Cheng}}, \ and\ \bibinfo {author}
  {\bibfnamefont {Y.}~\bibnamefont {Xu}},\ }\href {\doibase
  10.1016/j.ssc.2006.10.040} {\bibfield  {journal} {\bibinfo  {journal} {Solid
  State Commun.}\ }\textbf {\bibinfo {volume} {141}},\ \bibinfo {pages} {262}
  (\bibinfo {year} {2007})}\BibitemShut {NoStop}%
\bibitem [{\citenamefont {Palik}(1997)}]{Palik1997}%
  \BibitemOpen
  \bibfield  {author} {\bibinfo {author} {\bibfnamefont {E.~D.}\ \bibnamefont
  {Palik}},\ }\href@noop {} {\emph {\bibinfo {title} {{Handbook of Optical
  Constants of Solids}}}}\ (\bibinfo  {publisher} {Elsevier},\ \bibinfo
  {address} {New York},\ \bibinfo {year} {1997})\BibitemShut {NoStop}%
\end{thebibliography}
\end{document}